\documentclass[12pt]{iopart}

\usepackage{iopams}  
\expandafter\let\csname equation*\endcsname\relax
\expandafter\let\csname endequation*\endcsname\relax
\usepackage{amsmath,amssymb}
\usepackage{physics}
\usepackage{bm}
\usepackage{bbm}
\usepackage{hyperref}
\usepackage{color}
\usepackage[dvipsnames]{xcolor}
\usepackage[numbers,sort,compress]{natbib}
\hypersetup{
setpagesize=false,
 bookmarksnumbered=true,%
 bookmarksopen=true,%
 colorlinks=true,%
 linkcolor=Blue,
 citecolor=Blue,
 urlcolor = Blue,
}
\usepackage{graphicx, color}
\newcommand{\eq}[1]{Eq.~\eqref{#1}}
\newcommand{\p}{\partial}
\newcommand{\RM}[1]{\mathrm{#1}}
\newcommand{\Frak}[1]{\mathfrak{#1}}

\allowdisplaybreaks
\usepackage{etoolbox}

\makeatletter
\patchcmd{\section}
  {\normalsize}
  {\large\sffamily}
  {}
  {\PackageError{manuscript}{Failed to patch section size}{Check the definition of \string\section.}}

\patchcmd{\subsection}
  {\itshape}
  {\sffamily\upshape\bfseries}
  {}
  {\PackageError{manuscript}{Failed to patch subsection style}{Check the definition of \string\subsection.}}

\renewcommand{\tableofcontents}{%
  \section*{\contentsname}%
  \@starttoc{toc}%
}

\newcommand{\appendixtocformat}{%
  \patchcmd{\l@section}
    {\setlength\@tempdima{1.5em}}
    {\setlength\@tempdima{7.0em}}
    {}
    {\PackageWarning{manuscript}{Failed to patch appendix TOC}}%
  \renewcommand{\l@subsection}
    {\@dottedtocline{2}{1.5em}{2.5em}}%
  \renewcommand{\l@subsubsection}
    {\@dottedtocline{3}{3.8em}{7.5em}}%
}

\makeatother
\begin{document}

\title[ ]
{Scalar field theory for (chiral) active Brownian particles: bottom-up derivation revisited}

\author{Yuta Kuroda \& Thomas Speck}

\address{Institute for Theoretical Physics IV, University of Stuttgart, Heisenbergstr. 3, 70569 Stuttgart, Germany}
\ead{yuta.kuroda@itp4.uni-stuttgart.de}
\ead{thomas.speck@itp4.uni-stuttgart.de}
\vspace{10pt}
\begin{indented}
\item[]\today 
\end{indented}

\begin{abstract}
The active Brownian particle (ABP) model is one of the most widely used particle-based models in studies of scalar active fluids. 
A collection of interacting ABPs is known to exhibit intriguing phenomena that are absent in equilibrium counterparts. 
A prominent example is spontaneous phase separation, known as motility-induced phase separation (MIPS), which occurs even in the absence of explicit attractive interactions.
In parallel, the large-scale behavior of scalar active fluids is often modeled using continuum descriptions, namely, scalar field theories. 
Such scalar active field theories are usually formulated in a top-down manner, and establishing their connections to particle-based models remains challenging, with these connections not yet fully understood. 
To advance the microscopic derivation of scalar active field theories, we revisit the connection between ABPs interacting via a two-body potential in two dimensions and scalar field theories that include all possible terms up to fourth order in spatial gradients. 
Our approach is based on a pressure expansion and the renormalization group (RG) method in the context of singular perturbation theory. 
The RG method provides a systematic way to eliminate fast variables and identify the dynamics on the slow invariant manifold. 
Within this framework, scalar field theories can be obtained as RG flow equations. 
We also apply this method to a chiral variant of the ABP model, the chiral ABP (cABP) model, in which a constant torque biases particle rotation to the left or right. 
We show that the scalar field theory corresponding to the cABP model contains two ``odd'' terms in addition to those present in the scalar field theory for ABPs. 
The method developed here may also be useful for deriving hydrodynamic descriptions of other types of active matter systems. 
\end{abstract}

\tableofcontents

\section{Introduction}
Hydrodynamic descriptions and field theories are standard theoretical tools for analyzing the large-scale collective behavior of interacting many-body systems~\cite{Chaikin_Lubensky1995}.
A major advantage of such field-theoretic approaches is that they allow one to extract universal properties that do not depend on microscopic details.
These approaches have also been prevalently used to study systems driven out of equilibrium, such as active matter~\cite{Toner2005, Ramaswamy2010AnuualReview, Marchetti2013RMP}.
Although field equations are usually formulated based on the symmetries of the system, understanding how such large-scale descriptions can be derived from microscopic models is also a central concern in the statistical physics of both equilibrium~\cite{Irving_Kirkwood1950, Mori1958, McLennan1960, Sasa2014PRL, Hiura2023PRE} and nonequilibrium systems~\cite{Brey1998PRE, Bertini1997, Brader2012PRE}. 
For example, the celebrated Toner--Tu equation~\cite{Toner1995PRL, Toner1998PRE}, a coarse-grained description of polar active fluids, was originally formulated using a top-down approach based on symmetry considerations, and subsequent studies have attempted to derive coarse-grained equations from the corresponding microscopic particle-based models~\cite{Bertin2006PRE, Bertin2009JPhysA, Ihle2011PRE, Peshkov2014, Ihle2016JStatMech}.

Field theories are formulated in terms of slow variables, which typically correspond to conserved quantities or order parameters characterizing symmetry breaking.
In passive colloidal suspensions, particle number is the only conserved quantity, and their large-scale behavior can therefore be described solely by a scalar field related to the particle number density.
The field theory for a single conserved scalar variable is known as Model B and has been widely used to study phase separation in passive soft-matter systems~\cite{Hohenberg1977}.

The active counterparts of such systems are often referred to as scalar active fluids; examples include self-propelled colloids such as Janus particles~\cite{Volpe2011SoftMatter, Buttinoni2013PRL, Nisiguchi2015PRE}.
Active extensions of Model B have been proposed to study the collective behavior of such scalar active fluids~\cite{Wittkowski2014NatCom, Tjhung2018PRX, Cates2025}. 
Let $\varphi = \varphi(\bm r,t)$ be a conserved scalar field satisfying the continuity equation 
$\p_t \varphi = -\nabla\cdot \bm J$.
In scalar active fluids, the current $\bm J = \bm J(\bm r,t)$ is assumed to take the form
\begin{equation}
\bm J = - \Gamma \nabla\fdv{\mathcal F[\varphi]}{\varphi} 
+\bm J^\RM{act}.
\label{eq:AMB_current}
\end{equation}
The first term on the right-hand side corresponds to passive Model B and takes the form of an equilibrium contribution involving the (effective) free-energy functional $\mathcal F[\varphi]$, which takes the standard $\varphi^4$ form with a stiffness term, and $\Gamma$ denotes the effective mobility.
The second term, $\bm J^\RM{act}$, is an active current that cannot be expressed in terms of a free energy and therefore violates the fluctuation-dissipation relation or detailed balance.
Several forms of $\bm J^\RM{act}$ have been proposed.
The most widely used model, called active Model B+ (AMB+), includes the general form up to order $O(\nabla^3\varphi^2)$:
\begin{equation}
\bm J^\RM{act} = - \lambda \nabla |\nabla \varphi|^2 
+ \zeta (\nabla^2 \varphi) \nabla \varphi. 
\end{equation}
When $\lambda = \zeta = 0$, \eq{eq:AMB_current} reduces to passive Model B, which describes gradient flow dynamics.
Since nonzero $\lambda$- and $\zeta$-terms break time-reversal symmetry, AMB+ is known to modify the familiar properties of equilibrium phase separation~\cite{Nardini2017PRX, Tjhung2018PRX, Fausti2021PRL, Cates2023PRL, Fausti2024PRE, Hertag2026PRE}.
For example, the binodal densities are not determined by the two minima of the effective free energy, Ostwald ripening can be reversed, and microphase separation can occur~\cite{Tjhung2018PRX, Hertag2026PRE}.
AMB+ has also been used to study how liquid-gas criticality is modified in scalar active fluids~\cite{Caballero2018JStatMech, Papanikolaou2024, Fejos2026PRE, Bhowmick2026PRE}. 
In the recent work~\cite{Burekovic2026PRR}, a more general form of the active current $\bm J^\RM{act}$ has been proposed and studied. 

\begin{figure}[t]
\centering
\includegraphics[width=1.0\linewidth]{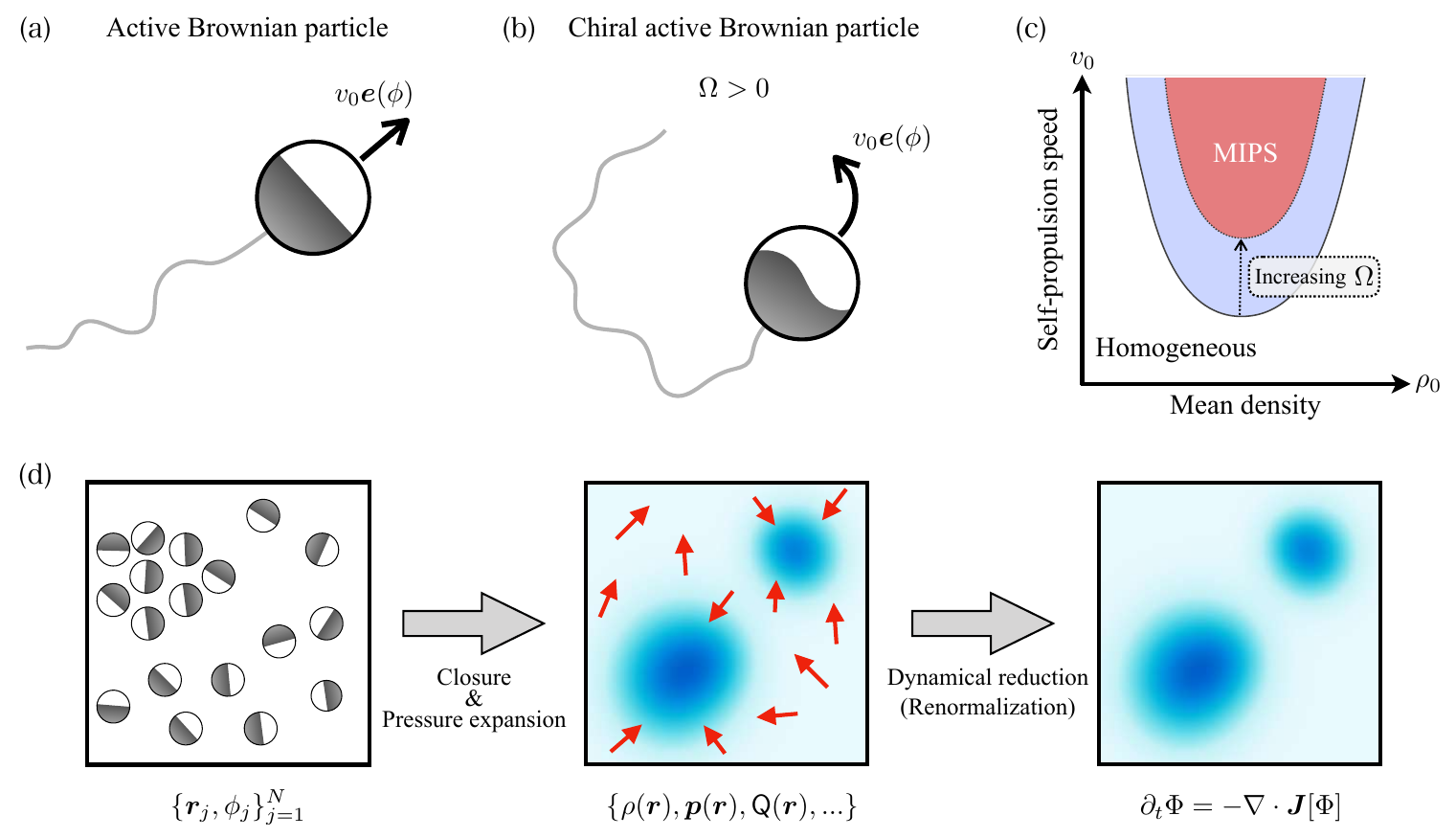}
\caption{\label{fig:model}Schematic illustration of (a)~the active Brownian particle model and (b)~the chiral active Brownian particle model.
(c)~Schematic of the phase diagram of fluid states of interacting (chiral) ABPs. The blue and red areas represent the coexistence regions in ABPs and cABPs, respectively. 
(d)~Coarse-graining procedure in this paper.   
}
\end{figure}

Particle-based models are also widely used to study scalar active fluids.
One of the most popular models is the active Brownian particle (ABP) model~\cite{Fily2012PRL, Bechinger2016RMP}.
In two dimensions, the position of the $j$-th particle, $\bm r_j$, evolves according to the overdamped Langevin equation
\begin{equation}
\dv{\bm r_j(t)}{t} = \mu\sum_{k(\neq j)}\bm f(\bm r_{jk}) + v_0 \bm e(\phi_j(t)) + \sqrt{2D_0} \bm \xi_j(t),
\label{eq:ABP_posi}
\end{equation}
where $\mu$ denotes the mobility, $\bm f(\bm r) = -\nabla U(r)$ is the interparticle force, with $U(r)$ denoting the pair potential, and $\bm r_{jk}=\bm r_j-\bm r_k$.
The second term, $v_0\bm e(\phi) = v_0(\cos\phi,\sin\phi)$, represents self-propulsion at a constant speed $v_0$. 
The last term represents noise from the surrounding medium.
$D_0$ denotes the translational diffusion coefficient, and $\bm \xi_j(t)$ is Gaussian white noise with zero mean and unit variance.
When $v_0$ is set to zero, Eq.~\eqref{eq:ABP_posi} reduces to passive Brownian dynamics at a temperature $D_0/\mu$.
The orientation angle $\phi_j(t)$ is assumed to undergo rotational diffusion according to
\begin{equation}
\dv{\phi_j(t)}{t} = \sqrt{\frac{2}{\tau_\RM{p}}}\eta_j(t), 
\label{eq:ABP_ori}
\end{equation}
where $\eta_j(t)$ is Gaussian white noise with zero mean and unit variance.
The timescale $\tau_\RM{p}$ is called the persistence time and characterizes the decay of orientational correlations through $\expval*{e_\alpha(\phi_j(t))e_\beta(\phi_k(t'))} \sim \exp(-|t-t'|/\tau_\RM{p})\delta_{\alpha,\beta}\delta_{j,k}$.
The persistent motion of an ABP is schematically shown in Figure~\ref{fig:model}(a).
Note that in the limit $\tau_\RM{p}\to 0$, with $D_\RM{act} = v_0^2\tau_\RM{p}/2$ held constant, this model also reduces to passive Brownian motion at an elevated effective temperature $T_\RM{eff} = (D_0 + D_\RM{act})/\mu$.
The most striking phenomenon observed in ABPs is spontaneous particle accumulation even in the absence of attractive interactions.
This phenomenon is called motility-induced phase separation (MIPS)~\cite{Cates2015AnnualReview} and has been extensively studied numerically~\cite{Fily2012PRL, Redner2013PRL, Stenhammar2014SoftMatt, Fily2014SoftMatt, Bialke2015PRL, Siebert2018PRE, Digregorio2018PRL}, theoretically~\cite{Taillerur2008PRL, Cates2013EPL, Bialk2013EPL, Solon2018PRE, Speck2021PRE}, and experimentally~\cite{Palacci2013Science, Buttinoni2013PRL, Linden2019PRL, Geyer2019PRX, Liu2019PRL}. 
The coexistence region of MIPS is known to resemble that of liquid-vapor phase separation in equilibrium systems~\cite{Siebert2018PRE, Digregorio2018PRL}, as schematically shown in Figure~\ref{fig:model}(c).
The nonequilibrium phenomena observed in ABPs are not limited to MIPS.
The ABP model has also been used to study, for example, spatial velocity correlations~\cite{Caprini2020PRL, Szamel2021EPL, Kuroda2023PRR}, slow glassy dynamics~\cite{Fily2014SoftMatt, Liluashvili2017PRE, Mandal2020NatComm, Janzen2022PRR, Paoluzzi2022CommPhys}, and two-dimensional crystallization~\cite{Bialke2012PRL, Digregorio2018PRL, Digregorio2022SoftMatter, Paliwal2020PRR, Shi2023PRL}.   

In the ABP model, particle number is the only conserved quantity, and there is no polar or nematic order.
One therefore expects its large-scale behavior to be described by a scalar field theory of the form given by Eq.~\eqref{eq:AMB_current}.
Indeed, several studies have shown that MIPS in ABPs shares phenomenological features with phase separation in AMB+~\cite{Shi2020PRL, Nakano2024PRR}.
Given these observations, it is natural to ask how, and in which asymptotic limit, AMB+ (and its generalizations) can be derived from ABPs.
Several studies have attempted to derive scalar field theories, including AMB+, from ABPs~\cite{Speck2014PRL, Speck2015JCP, Speck2022PRE, Vrugt2023J.Phys.Cond.Matt, Pinto-Goldberg2026}.
For ABPs interacting via a pairwise potential, the coarse-graining procedure essentially involves two issues:
(i)~how to truncate the BBGKY-type hierarchy, and (ii)~how to eliminate fast field variables, such as polarization.
For the first issue, a standard approach is based on a pressure expansion~\cite{Bickmann2020, Speck2022PRE, Vrugt2023J.Phys.Cond.Matt}.
An alternative approach is based on a Boltzmann-like kinetic theory, which is applicable only in the dilute limit for hard-core particles~\cite{Pinto-Goldberg2026}.
For the second issue, the adiabatic or quasi-stationary approximation, which simply sets the time derivatives of the fast variables to zero, is often employed~\cite{Cates2013EPL, Bialk2013EPL}.
To go beyond the adiabatic approximation, several more sophisticated methods for eliminating fast variables, such as weakly nonlinear analysis~\cite{Speck2014PRL, Rapp2019EPJE} and eigenmode expansion~\cite{Speck2022PRE}, have been applied.
Among the many studies aiming to derive field theories for interacting ABPs, those by Speck~\cite{Speck2022PRE} and te Vrugt {\it et al.}~\cite{Vrugt2023J.Phys.Cond.Matt} have yielded scalar field theories with the same form as AMB+.
However, Speck~\cite{Speck2022PRE} employed a phenomenological pressure expansion close to the critical point, and the resulting field-theory parameters therefore cannot be determined from the microscopic parameters.
While the pressure expansion carried out by te Vrugt {\it et al.}~\cite{Vrugt2023J.Phys.Cond.Matt} contains no phenomenological parameters, they used the adiabatic approximation to eliminate fast field variables, which generally misses several relevant contributions, as recently demonstrated in Ref.~\cite{Burekovic2026PRR}. 
Burekovic {\it et al.}~\cite{Burekovic2026PRR} have also derived a scalar field theory for active particles with quorum-sensing interactions.

In this paper, we revisit the microscopic derivation of scalar field theories up to fourth order in spatial gradients for ABPs with pairwise steric interactions.
Our derivation is based on the pressure expansion~\cite{Speck2022PRE, Vrugt2023J.Phys.Cond.Matt} and the renormalization group (RG) method in the context of singular perturbation theory~\cite{Chen1994PRL, Kunihiro2022}.
Following the standard procedure~\cite{Bialk2013EPL, Speck2015JCP}, we start from the exact equation for the one-body distribution function corresponding to Eqs.~\eqref{eq:ABP_posi} and \eqref{eq:ABP_ori}, which contains an interaction term corresponding to a ``pressure'' gradient.
To close this equation, we make several assumptions about the pair distribution functions~\cite{Bialk2013EPL}.
We then expand the interaction term in spatial gradients and orientational moments to obtain an expression for the pressure gradient up to fourth order in spatial gradients, $O(\nabla^4)$.
Using this expression for the pressure gradient, we construct effective hydrodynamic equations that contain all the terms relevant to deriving an $O(\nabla^4)$ field theory within the closure approximation.
To obtain a scalar field theory, we need to eliminate fast variables.
To this end, we employ the RG method in the context of singular perturbation theory, which allows us to systematically reduce the dynamics to those on a slow invariant manifold~\cite{Kunihiro2022}.
The resulting field equation has the same form as that introduced in Ref.~\cite{Burekovic2026PRR}. 
Near the critical point, the field equation reduces to AMB+.
In principle, all the coefficients in the equations can be determined from the interaction potential and the pair distribution function.
Figure~\ref{fig:model}~(d) outlines the steps in our derivation of the scalar field theory.

We further study a chiral variant of the ABP model, known as the chiral active Brownian particle (cABP) model, in which the motion of each particle is biased to the left or right by a constant torque~\cite{Teeffelen2008PRE, Ma2017SoftMat}, as depicted in Figure~\ref{fig:model}(b).
This model has been widely used to study the effects of chirality on the collective behavior of self-propelled particles~\cite{Lei2019SciAdv, Ma2022JCP, Debets2023PRL, Kuroda2023JStatMech, Caprini2024CommPhys, Kuroda2025PRR, Kuroda2025PRE, Jeong2025PRR}.
The effects of chirality on MIPS have also been investigated.
It has been shown that MIPS tends to be suppressed as chirality increases [Figure~\ref{fig:model}(c)], and microphase separation can also occur~\cite{Ma2022JCP, Bickmann2022JCP}.
Several studies have considered field theories for cABPs~\cite{Ma2022JCP, Bickmann2022JCP, Sansa2022JCP, Kuroda2023JStatMech, Kalz2024JPhysA}.
In particular, Kalz {\it et al.}~\cite{Kalz2024JPhysA} have derived AMB+ from the cABP model with hard-core interactions using a kinetic approach and the adiabatic approximation.
Recently, a scalar field theory for quorum-sensing chiral particles was reported in Ref.~\cite{Burekovic2026}.
Here, we show that the $O(\nabla^4)$ field theory for cABPs contains two ``odd'' nonequilibrium contributions in addition to those present in the density current of their nonchiral counterparts. 

This paper is organized as follows.
In Section~\ref{sec:framework}, we describe the setup and outline the basic steps for the derivation of scalar field theories.
We also introduce several assumptions to obtain a closed equation for the one-body distribution function.
In Section~\ref{sec:pressure}, we then develop a pressure expansion within the closure approximation that includes the terms necessary to obtain an $O(\nabla^4)$ theory.
Section~\ref{sec:hydrodynamics} summarizes the effective hydrodynamic equations, which are coupled equations for the density, polarization, and nematic tensor.
Section~\ref{sec:RG} forms the core of this paper.
To eliminate the fast degrees of freedom, we apply the RG method and obtain the slow dynamics of a scalar field.
We also identify the invariant manifold and explain why the naive adiabatic approximation should not be used to derive an $O(\nabla^4)$ theory.
In Section~\ref{sec:chiral}, we apply the same pressure expansion and the RG method to interacting cABPs and derive an $O(\nabla^4)$ theory that contains odd terms.
Finally, Section~\ref{sec:conclusions} concludes the paper.

\section{Framework}
\label{sec:framework}

\subsection{Setup}
We consider $N$ interacting particles that obey Eqs.~\eqref{eq:ABP_posi} and \eqref{eq:ABP_ori}.
The interaction potential $U(r)$ is assumed to be short-ranged: there is a cut-off length $r_\RM{c}$ such that $U(r) = 0$ for $r > r_\RM{c}$.
Our starting point is the one-body distribution function defined by
\begin{equation}
\Psi(\bm r,\phi,t) := \expval{\sum_{j=1}^{N} \delta(\bm r-\bm r_j(t)) \delta_{2\pi}(\phi- \phi_j(t))},
\end{equation}
where $\delta_{2\pi}(\phi) = \sum_{n\in \mathbb Z}\delta(\phi + 2n\pi)$.
Taking the time derivative of this definition and using Eqs.~\eqref{eq:ABP_posi} and \eqref{eq:ABP_ori} yields
\begin{equation}
\p_t \Psi =  -\nabla\cdot \qty[
\mu \bm G  + v_0\bm e(\phi) - D_0\nabla] \Psi
+\frac{1}{\tau_\RM{p}} \p_\phi ^2\Psi.
\label{eq:one-body1}
\end{equation}
Here, $\bm G = \bm G(\bm r,\phi,t)$ denotes the conditional force arising from pair interactions:
\begin{equation}
\bm G(\bm r,\phi,t) 
:= \int_{V}\dd[2]\bm r'\int_{-\pi}^{\pi}\dd\phi' \, \bm f(\bm r-\bm r')\Psi_2(\bm r',\phi'|\bm r,\phi;t ),
\label{eq:con_force_def}
\end{equation}
where $\Psi_2(\bm r',\phi'|\bm r,\phi;t ) := \Psi_2(\bm r,\phi; \bm r',\phi';t )/\Psi(\bm r, \phi, t)$ is the conditional two-body distribution function, and $\Psi_2(\bm r,\phi; \bm r',\phi';t )$ is defined by
\begin{equation}
\Psi_2(\bm r,\phi; \bm r',\phi';t ):= \expval{\sum_{j\neq k} \delta(\bm r-\bm r_j(t)) \delta_{2\pi}(\phi- \phi_j(t))\delta(\bm r'-\bm r_k(t)) \delta_{2\pi}(\phi'- \phi_k(t))}.
\end{equation}
The summation $\sum_{j\neq k}$ runs over all $j,k\in\{1,2,...,N\}$ except for $j = k$.

Our goal is to obtain a closed equation for a scalar field describing the slow, large-scale dynamics. 
This slow variable is related to the number density defined by
\begin{equation}
\rho(\bm r,t)
= \expval{\sum_{j=1}^{N}\delta(\bm r-\bm r_j(t))}
= \int_{-\pi}^{\pi}\dd\phi \ \Psi(\bm r,\phi,t). \label{eq:rho_def}
\end{equation}
This derivation involves three main steps:
\begin{itemize}
    \item[(i)]{\it Closure of the BBGKY-type hierarchy:} The interaction term $\bm G$ in Eq.~\eqref{eq:one-body1} contains the two-body distribution $\Psi_2$. 
    Hence, Eq.~\eqref{eq:one-body1} is not closed in terms of $\Psi$. 
    The equation for $\Psi_2$, in turn, contains the three-body distribution function. 
    An approximation is therefore required to close this hierarchy.
    \item[(ii)]{\it Pressure expansion:} After closing the hierarchy in some way, the interaction term $\bm G$, which is related to the pressure gradient, is expressed in terms of the static pair distribution function. 
    We then expand $\bm G$ in terms of the field variables and their gradients.  
    \item[(iii)]{\it Elimination of fast degrees of freedom:} After the preceding two steps, one obtains an equation for the density field $\rho(\bm r,t)$ that includes other field variables, such as the polarization and nematic tensor. These variables decay in the long-time limit.
    They must therefore be eliminated asymptotically.
\end{itemize}
For step~(i), we employ the approximation for the pair distribution function proposed in Ref.~\cite{Bialk2013EPL}.
In Section~\ref{subsec:closure}, we describe this closure approximation.
This closure is crude and constitutes the only nonsystematic approximation in our derivation.
Step~(ii) can be carried out once we adopt the closure approximation, as shown in Section~\ref{sec:pressure}.
Step~(iii) is the core of this paper.
We carry out this step using the renormalization group method in Section~\ref{sec:RG}.
We emphasize that such a singular perturbation method is necessary to obtain the correct dynamics on the slow invariant manifold.

\subsection{Closure approximation}
\label{subsec:closure}
Here, we introduce an approximation to close Eq.~\eqref{eq:one-body1} in terms of $\Psi$ alone.
We first introduce the pair distribution function through
\begin{equation}
\Psi_2(\bm r'|\bm r,\phi;t ) = g(\bm r',\bm r, \phi; t)\rho(\bm r',t),
\end{equation}
where $\Psi_2(\bm r'|\bm r,\phi;t ) = \int_{-\pi}^\pi\dd\phi'\, \Psi_2(\bm r',\phi'|\bm r,\phi;t )$.
This is simply the definition of $g(\bm r',\bm r, \phi; t)$.
We make the following three assumptions regarding $g(\bm r',\bm r, \phi; t)$~\cite{Bialk2013EPL}:
\begin{itemize}
    \item The time dependence of the pair distribution function is negligible:
    \begin{equation}
    g(\bm r',\bm r, \phi; t) \simeq g(\bm r',\bm r, \phi).
    \label{eq:assum_gr_1}
    \end{equation}
    This assumption amounts to identifying the time-dependent pair distribution function with its steady-state counterpart, in analogy with the adiabatic approximation used in dynamical density functional theory~\cite{Vrugt02042DDFT}.
    \item The system is homogeneous, i.e., the pair distribution function is independent of the choice of spatial origin:
    \begin{equation}
    g(\bm r',\bm r, \phi) \simeq g(\bm r' - \bm r, \phi).
    \label{eq:assum_gr_2}
    \end{equation}
    \item The pair distribution function is axisymmetric
    with respect to the orientational vector $\bm e(\phi)$:
    \begin{equation}
    g(\bm r,\phi) = \tilde g(r,\theta - \phi)
    \ \ \ \ \ \ \mbox{and}\ \ \ \ \ \ 
    \tilde g(\cdot,\phi) = \tilde g(\cdot,-\phi).
    \label{eq:assum_gr_3}
    \end{equation}
    Here, $r = |\bm r|$, and $\theta$ is the angle of the vector $\bm r$ with respect to the $x$-axis.
\end{itemize}
Although the first assumption is often used in the context of active matter, this is crude and is difficult to justify. 
The second assumption is reasonable when the system remains homogeneous or the density variations occur over sufficiently long lengths. 
However, the validity of this approximation remains unclear, especially when the system undergoes phase separation.
The third assumption is based on numerical observations of the pair distribution function in the steady state~\cite{Bialk2013EPL, Poncet2021PRE}. 
Numerical simulations in homogeneous ABP systems have shown that the pair distribution function is axisymmetric with respect to $\bm e(\phi)$ and takes larger values in the forward direction. 
However, this assumption is somewhat subtle for the cABP model~\cite{Kuroda2026} (see also \ref{ape:gr}). 

These assumptions about the pair distribution function are the only nonsystematic approximation in our derivation.
The conditional force is then approximated as
\begin{align}
\bm G(\bm r,\phi,t) 
&\simeq  \int_{V}\dd[2]\bm r' \ \bm f(\bm r-\bm r')g(\bm r'-\bm r, \phi)\rho(\bm r',t) \notag \\
&= \int_{r' < r_\RM{c}}\dd[2]\bm r'\ [\nabla' U(r')]
g(\bm r', \phi)\rho(\bm r + \bm r',t).
\end{align}
Here, $r_\RM{c}$ is the cutoff length scale of the potential $U(r)$.

\subsection{Rescaling of length and time}
\label{subsec:scaling}
To systematically evaluate which terms are relevant for $O(\nabla^4)$ theories in the hydrodynamic scaling regime, we introduce a small parameter involving spatial gradients.
We first introduce the macroscopic length and time scales $\ell_\RM{macro}$ and $\tau_\RM{macro}$, which correspond, for example, to the characteristic length and time scales of an instability.
We then rescale time and position as $ t \to \tau_\RM{macro} t$ and $\bm r \to \ell_\RM{macro} \bm r $.
We also define the two dimensionless parameters
\begin{equation}
\epsilon_\ell := \frac{\ell_\RM{p}}{\ell_\RM{macro}},
\ \ \ \ \ \ \ 
\epsilon_\tau := \frac{\tau_\RM{p}}{\tau_\RM{macro}},
\end{equation}
where $\tau_\RM{p}$ and $\ell_\RM{p} = v_0\tau_\RM{p}$ are the characteristic microscopic time and length scales.
We rewrite Eq.~\eqref{eq:one-body1} in terms of the rescaled time and position.
We impose the scale-separation condition $\epsilon_\ell, \epsilon_\tau \ll 1$.
In particular, we consider the limit $\epsilon_\ell, \epsilon_\tau \to 0$ while requiring the diffusion term to remain $O(1)$, i.e., $\epsilon_\ell^2/\epsilon_\tau = O(1)$. 
Since constant factors can be absorbed into the definitions of $\tau_\RM{macro}$ and $\ell_\RM{macro}$, one can set $\epsilon := \epsilon_\ell = \sqrt{\epsilon_\tau}$.
We thus obtain
\begin{equation}
\p_t \Psi(\bm r,\phi,t) =  -\nabla\cdot \qty[
 \frac{1}{\epsilon} \frac{\mu\tau_\RM{p}}{\ell_\RM{p}} \bm G(\bm r,\phi,t)   + \frac{1}{\epsilon}\bm e(\phi) -  d_0\nabla] \Psi(\bm r,\phi,t) 
+ \frac{1}{\epsilon^2}   \p_\phi ^2\Psi(\bm r,\phi,t),
\label{eq:one_body_rescaled}
\end{equation}
with dimensionless ratio $d_0:=D_0\tau_\RM{p}/\ell_\RM{p}^2=D_0/(v_0^2\tau_\RM{p})$ of passive to active diffusion.
We further introduce the fast time variable  
\begin{equation}
\Frak t := \frac{t}{\epsilon^2}
\label{eq:fast_time}
\end{equation}
and rewrite the equation for the one-body distribution as
\begin{equation}
\p_{\Frak t} \Psi(\bm r,\phi, \Frak t) =  -\nabla\cdot \qty[
\bm {\mathcal G}(\bm r,\phi,\Frak t)   + \epsilon\bm e(\phi) - \epsilon^2 d_0\nabla] \Psi(\bm r,\phi,\Frak t) 
+ \p_\phi ^2\Psi(\bm r,\phi, \Frak t), 
\label{eq:one-body2}
\end{equation}
where 
\begin{equation}
\bm {\mathcal G}(\bm r,\phi, \Frak t) := \epsilon \frac{\mu \tau_\RM{p}}{\ell_\RM{p}}\bm G (\bm r,\phi,t). 
\label{eq:con_force_scaled}
\end{equation}

\section{Pressure expansion}
\label{sec:pressure}

Our aim is to derive an equation for a scalar field related to the density.
Using the definition of the density, Eq.~\eqref{eq:rho_def}, integrating Eq.~\eqref{eq:one-body2} over $\phi$ yields the continuity equation $\p_{\Frak t}\rho = -\nabla\cdot \bm J^\rho$, where the density current $\bm J^\rho =\bm J^\rho (\bm r,\Frak t)$ is given by
\begin{align}
\bm J^\rho = \epsilon \bm p - \epsilon ^2 d_0 \nabla\rho - \nabla\cdot {\mathsf \Pi}^\rho,
\label{eq:current1}
\end{align}
where 
\begin{equation}
\bm p (\bm r,\Frak t) 
= \int_{-\pi}^{\pi}\dd\phi \ \bm e(\phi)\Psi(\bm r,\phi,\Frak t)
\end{equation}
is the polarization, and
\begin{align}
 \nabla\cdot{\mathsf \Pi}^\rho(\bm r,\Frak t) := -\int_{-\pi}^{\pi}\dd\phi\ \bm {\mathcal G}(\bm r,\phi,\Frak t)\Psi(\bm r,\phi,\Frak t).
\end{align}
The tensor ${\mathsf \Pi}^\rho$ is the interaction contribution to the pressure tensor.
The explicit form of ${\mathsf \Pi}^\rho$ is given by the Irving--Kirkwood formula~\cite{Irving_Kirkwood1950, Kruger2018JCP}.
Specifically, we aim to find an equation for a scalar field up to $O(\nabla^4)$.
Because the small parameter $\epsilon$ is accompanied by a spatial gradient in our framework, our next task is to find an expression for the conditional force $\bm {\mathcal G}$ that yields all contributions to the pressure gradient up to $O(\epsilon^4\nabla^3)$.

\subsection{Order estimation} 
\label{subsec:order_est}
In preparation for the pressure expansion, we first consider the rank-$n$ tensor
\begin{align}
\mathsf Q_n(\bm r, \Frak t) := \int_{-\pi}^{\pi}\dd\phi\ \qty[\bm e^{\otimes n}(\phi)]^\RM{ST}\Psi(\bm r,\phi, \Frak t)
\label{eq:ori_moment}
\end{align}
for $n\geq 1$ so that $\bm p = \mathsf Q_1 $ is the polarization and $\mathsf Q:= \mathsf Q_2 $ is the nematic tensor.
$\mathsf T^\RM{ST}$ denotes the traceless symmetric part of the tensor $\mathsf T$.
For example, when $\mathsf T$ is a second-rank tensor, $T^\RM{ST}_{\alpha,\beta} = (T_{\alpha,\beta} + T_{\beta,\alpha})/2 - T_{\mu,\mu}\delta_{\alpha,\beta}/2$, where repeated indices are summed over. 
For a tensor of rank greater than $2$, we define a trace as a contraction over two indices.
A symmetric rank-$n$ tensor $\mathsf T$ is traceless if every contraction over two indices yields the zero tensor of rank $(n-2)$: $\delta_{\alpha_j,\alpha_k}T_{\alpha_1,\alpha_2,...,\alpha_n} =0$ for all $j < k$.

As shown below, the conditional force can be expanded in terms of $[\bm e^{\otimes n}(\phi)]^\RM{ST}$ and spatial gradients.
We therefore first determine the order of $\mathsf Q_n(\bm r, \Frak t)$ in $\epsilon$.

To estimate the order of $\mathsf Q_n(\bm r, \Frak t)$, we expand $\Psi(\bm r,\phi, \Frak t)$ in a Fourier series:
\begin{align}
\Psi(\bm r,\phi, \Frak t) = \frac{1}{2\pi}\sum_{n\in \mathbb Z}\psi_n (\bm r, \Frak t)e^{-in\phi}
\end{align}
with
\begin{equation}
\psi_n (\bm r, \Frak t) =  \int_{-\pi}^{\pi}\dd\phi\ \Psi(\bm r,\phi, \Frak t)e^{in\phi}.
\end{equation}
Because the components of $[\bm e^{\otimes n}(\phi)]^\RM{ST}$ are proportional to either $\sin(n\phi)$ or $\cos(n\phi)$, every component of $\mathsf Q_n$ is proportional to either $\RM{Re}[\psi_\abs{n}]$ or $\RM{Im}[\psi_\abs{n}]$.
In Fourier space, Eq.~\eqref{eq:one-body2} becomes
\begin{align}
\p_{\mathfrak t}\psi_{n} = -\epsilon \p_z \psi_{n+1} -\epsilon \p_{z^*} \psi_{n-1} + \epsilon^2 d_0 \nabla^2 \psi_n - n^2 \psi_n - I_n,
\end{align}
where $\p_z  = (\p_x - i\p_y)/2$ and the interaction term is defined by
\begin{align}
 I_n(\bm r) := \int_{-\pi}^{\pi}\dd\phi\ \nabla\cdot \qty[\bm{\mathcal G}(\bm r,\phi)\Psi(\bm r,\phi)]e^{in\phi}.
 \label{eq:interaction_Fourier}
\end{align}
For simplicity, we first consider the noninteracting case, $I_n(\bm r)=0$.
We now expand $\psi_{n}$ as
\begin{align}
\psi_{n} = \psi_{n} ^{(0)} + \epsilon \psi_{n} ^{(1)} +  \epsilon^2 \psi_{n} ^{(2)} +\cdots .
\end{align}
We consider the long-time regime $\Frak t \gg 1$, neglecting exponentially decaying terms.
At zeroth order, we have $\psi_{n} ^{(0)}(\bm r,\mathfrak t) =\Phi(\bm r)$ for $n=0$ and $\psi_{n} ^{(0)}(\bm r,\mathfrak t) = 0$ otherwise, where $\Phi(\bm r)$ is time independent.
One can also show that $\psi_2^{(1)}=0$ and $\psi_3^{(1)}=0$.
In general, $\psi_n^{(m)}=0$ for $m<|n|$.
The equation for $n=1$ at $O(\epsilon ^1)$ gives
\begin{equation}
 \psi_{1} ^{(1)}= - \p_{z^*}\psi^{(0)}_0 =O(\nabla\Phi).
\end{equation}
Therefore, $ \psi_{1}$ scales as $ \psi_{1} = O(\epsilon\nabla)$ at leading order for $\Phi = O(1)$.
The equation for $n=2$ at $O(\epsilon ^2)$ gives
\begin{equation}
 \psi_{2} ^{(2)}= - \frac{1}{4}\p_{z^*}\psi^{(1)}_1.
\end{equation}
Therefore, $ \psi_{2}$ scales as $ \psi_{2} = O(\epsilon^2\nabla^2)$ at leading order.
Continuing this argument, we find inductively that $\psi_n = O\qty(\epsilon^{\abs{n}}\nabla^{\abs{n}})$.
As shown below, the interaction term satisfies $I_n = O\qty(\epsilon^{\abs{n}}\nabla^{\abs{n}})$, and $I_n$ therefore does not alter the order of $\psi_n$.
Because every component of $\mathsf Q_n$ is proportional to either $\RM{Re}[\psi_\abs{n}]$ or $\RM{Im}[\psi_\abs{n}]$, we estimate the leading-order scaling of $\mathsf Q_n$ as
\begin{align}
\mathsf Q_n = O\qty(\epsilon^n\nabla^{n}).
\label{eq:lowest_order}
\end{align}

\subsection{Gradient expansion}

The scaled conditional force defined in Eq.~\eqref{eq:con_force_scaled} is given by
\begin{align}
\bm {\mathcal G}(\bm r,\phi) \simeq \mu \tau_\RM{p}\int_{s<r_\RM{c}/\ell_\RM{macro}}\dd[2]\bm s\ \frac{\bm s}{s}U'(s)g(\bm s,\phi)\rho (\bm r + \bm s).
\end{align}
Here, we have used the assumptions in Eqs.~\eqref{eq:assum_gr_1} and \eqref{eq:assum_gr_2}.
Assuming that the cutoff length of the potential satisfies $r_\RM{c} = O(\ell_\RM{p})$ so that $r_\RM{c}/\ell_\RM{macro} = O(\epsilon)$, we expand the density in the integral as
\begin{equation}
\rho(\bm r + \bm s) = \rho(\bm r) + s_\beta \p_\beta\rho(\bm r) +\frac{1}{2} s_\beta s_\mu \p_\beta\p_\mu\rho(\bm r)  + \frac{1}{3!} s_\beta s_\mu s_\nu \p_\beta\p_\mu \p_\nu \rho(\bm r) + O\qty( \epsilon^4 \nabla^4).
\end{equation}
Note that $s$ in the integral is $O(\epsilon)$ and $\int_{s<r_\RM{c}/\ell_\RM{macro}}\dd[2]\bm s\ U'(s)g(\bm s,\phi) = O(\epsilon)$.
Thus, the conditional force is expanded as 
\begin{equation}
\begin{split}
\mathcal G_\alpha(\bm r,\phi) =
\epsilon g^{(0)}_\alpha(\phi)\rho(\bm r) 
+ \epsilon^2 g_{\alpha,\beta}^{(1)}(\phi)\p_\beta\rho(\bm r)
+\epsilon^3 g_{\alpha,\beta,\mu}^{(2)}(\phi)\p_\beta\p_\mu\rho(\bm r) \\
+\epsilon^4 g_{\alpha,\beta,\mu, \nu}^{(3)}(\phi)\p_\beta\p_\mu\p_\nu\rho(\bm r) 
+ O(\epsilon^5\nabla^4).
\label{eq:gradient_expansion}
\end{split}
\end{equation}
Each coefficient is $O(1)$ and is defined by an integral involving the pair distribution function and the pairwise potential (see \ref{ape:pressure_expansion} for the definitions).
In \ref{ape:pressure_expansion}, we explicitly calculate each coefficient under the assumption that $g(\bm r,\phi)$ is axisymmetric, as stated in Eq.~\eqref{eq:assum_gr_3}.
Neglecting contributions of $O(\epsilon^5\nabla^4)$ and higher to the pressure gradient, we approximate the conditional force as
\begin{equation}
\begin{split}
\bm{\mathcal G}(\bm r,\phi) \simeq 
&- \epsilon\gamma \rho(\bm r)\bm e(\phi) -  \epsilon^2[B_0\mathbbm 1 + B_2 \mathsf q(\phi)]\nabla\rho(\bm r) 
\\
&- \epsilon^3C\qty[\bm e(\phi)\nabla^2 + 2\bm e(\phi)\cdot \nabla\nabla]\rho(\bm r)
+ \epsilon^4K \nabla\nabla^2\rho(\bm r),
\label{eq:con_force_expansion}
\end{split}
\end{equation}
where $\mathsf q(\phi) = [\bm e(\phi)\bm e(\phi)]^\RM{ST}$.
The coefficients $\gamma$, $B_0$, $B_2$, $C$, and $K$ are determined by the pair distribution function and the interaction potential. 
The pressure gradient is thus given by
\begin{equation}
\begin{split}
\nabla \cdot \mathsf \Pi^\rho(\bm r) 
=
&\epsilon\gamma\rho\bm p + \epsilon^2\qty[B_0\rho\mathbbm 1 + B_2\mathsf Q]\nabla\rho
\\
&+\epsilon^3C\qty[\bm p\nabla^2 + 2\bm p\cdot\nabla\nabla]\rho 
- \epsilon^4 K \rho\nabla\nabla^2\rho + O(\epsilon^5\nabla^4),
\label{eq:pressure_expansion}
\end{split}
\end{equation}
where
$\mathsf Q = \mathsf Q_2$
is the nematic tensor [see Eq.~\eqref{eq:ori_moment}].
In this expression, the terms proportional to $\gamma$, $B_2$, and $C$ are active contributions and should therefore all vanish in the limit $v_0\to 0$ or $\tau_\RM{p}\to 0$.
Equation~\eqref{eq:con_force_expansion} also implies that the interaction term in Eq.~\eqref{eq:interaction_Fourier} is $O(\epsilon^n\nabla^n)$ at leading order.

\section{Effective hydrodynamic equations}
\label{sec:hydrodynamics}

The pressure gradient in Eq.~\eqref{eq:pressure_expansion} contains the density, polarization, and nematic tensor.
Here, we summarize the coupled equations for these three field variables, retaining the terms required to derive an $O(\nabla^4)$ theory for the scalar field.

For convenience, we introduce a reference density and work with dimensionless field variables.
Substituting Eq.~\eqref{eq:pressure_expansion} into the density current $\bm J^\rho$ in Eq.~\eqref{eq:current1},
we find that the $\gamma$-term yields the density-dependent speed
\begin{equation}
u(\rho) := 1- \gamma \rho  = 1- \frac{\rho}{\rho_{*}},
\label{eq:effective_speed}
\end{equation}
where $\rho_{*} := 1/\gamma$ is the density at which the effective self-propulsion speed vanishes. 
We nondimensionalize the one-body distribution function by dividing $\Psi$ by the reference density $\rho_*$. 
The field variables $\rho$, $\bm p$, and $\mathsf Q$ are then dimensionless as well.
Equation~\eqref{eq:effective_speed} then becomes $u(\rho) = 1- \rho$, and the expanded conditional force given by Eq.~\eqref{eq:con_force_expansion} is characterized by the dimensionless coefficients
\begin{equation}
b_0 := \rho_{*}B_0,\ \ \ 
b_2 := \rho_{*}B_2,\ \ \ 
c:= \rho_{*}C,\ \ \ 
k:= \rho_{*}K.
\end{equation}

With this nondimensionalization, the density equation reads $\p_{\Frak t}\rho = -\nabla\cdot \bm J^\rho$, with the components of the density current given by
\begin{equation}
\begin{split}
J^\rho_\alpha =
 \epsilon u(\rho)p_\alpha 
 - \epsilon^2 d(\rho)\p_\alpha\rho 
- \epsilon^2 b_2 Q_{\alpha,\beta} \p_\beta\rho 
 - \epsilon^3 c[p_\alpha \nabla^2 
 + 2p_\beta  \p_\beta \p_\alpha]\rho 
 \\
 + \epsilon^4 k \rho \p_\alpha\nabla^2\rho
 +O(\epsilon^5\nabla^4),
 \label{eq:density_current}
\end{split}
\end{equation}
where we define
\begin{equation}
d(\rho) := d_0 + b_0\rho.
\end{equation}
To construct the equation for the polarization, we recall that $\bm p = O(\epsilon \nabla)$, as discussed in Section~\ref{subsec:order_est}, and that we retain the terms up to $O(\epsilon^4\nabla^3)$ in the density current.
Because the density current contains a term proportional to $\epsilon \bm p$, terms up to $O(\epsilon^3\nabla^3)$ must be retained in the evolution equation for $\bm p$. 
The evolution equation for $\bm p = \bm p(\bm r,\Frak t)$ is
\begin{equation}
\p_{\Frak t}p_\alpha = - p_\alpha
-\epsilon \p_\beta\qty[Q_{\alpha,\beta} + \frac{\rho}{2}\delta_{\alpha,\beta}] + \epsilon^2 d_0 \nabla^2p_\alpha -\p_\beta \Pi^\RM{p}_{\alpha,\beta},
\label{eq:polarization}
\end{equation}
where the interaction contribution $\Pi^\RM{p}_{\alpha,\beta} = \Pi^\RM{p}_{\alpha,\beta}(\bm r,\Frak t)$ is given by
\begin{align}
\Pi^\RM{p}_{\alpha,\beta}(\bm r,\Frak t):= \int_{-\pi}^{\pi}\dd\phi\ \qty[e_\alpha(\phi)\mathcal G_\beta(\bm r,\phi, \Frak t)]\Psi(\bm r,\phi,\Frak t).
\end{align}
Substituting Eq.~\eqref{eq:con_force_expansion} into this equation and retaining terms up to $O(\epsilon^3\nabla^2)$ gives
\begin{equation}
\begin{split}
\Pi^\RM{p}_{\alpha,\beta} = 
- \epsilon \rho\qty[Q_{\alpha,\beta} + \frac{\rho}{2}\delta_{\alpha,\beta} ]
-\epsilon^2 b_0 p_\alpha \p_\beta\rho 
- \frac{\epsilon^2 b_2}{2} \qty( \frac{1}{2}\delta_{\alpha,\beta}p_\mu \p_\mu\rho -[p_\alpha\p_\beta\rho]^\RM{A} )
\\
-\epsilon^3 c \rho\qty[\frac{1}{2}\nabla^2\delta_{\alpha,\beta} +\p_\alpha \p_\beta]\rho
+ O(\epsilon^4\nabla^3),
\end{split}
\end{equation}
where $T_{\alpha,\beta}^\RM{A}:= (T_{\alpha,\beta} - T_{\beta,\alpha})/2$ is the antisymmetric part of the tensor $\mathsf T$.
We also require an equation for the nematic tensor. 
The density current contains a term proportional to $\epsilon^2\mathsf Q\nabla\rho$, and $\mathsf Q = O(\epsilon^2\nabla^2)$.
We therefore retain only the leading-order terms of $O(\epsilon^2\nabla^2)$ in the equation for the nematic tensor:
\begin{equation}
\p_{\Frak t}Q_{\alpha,\beta} = - 4 Q_{\alpha,\beta} -\frac{\epsilon}{2}\qty[\p_\alpha ( u(\rho)p_\beta )]^\RM{ST} + \frac{\epsilon^2 b_2}{4}\qty[\p_\alpha (\rho\p_\beta\rho)]^\RM{ST} + O(\epsilon^4\nabla^4).
\label{eq:nematic}
\end{equation}
Higher moments $\mathsf Q_n$ for $n\geq 3$ can be neglected because they do not contribute to the derivation of an $O(\nabla^4)$ theory.

\section{Dynamical reduction}
\label{sec:RG}

The equations for the polarization and nematic tensor derived above contain damping terms, and these fields therefore relax to values determined by the density alone.
To describe the dynamics solely in terms of the density field, we eliminate these fast fields.
One of the simplest and most widely used methods is the adiabatic or quasi-stationary approximation, which naively eliminates all fast fields by setting their time derivatives to zero~\cite{Bialk2013EPL, Vrugt2023J.Phys.Cond.Matt}. 
However, this method is known to yield unphysical reduced equations~\cite{Burekovic2026PRR}. 

To avoid this issue and carry out dynamical reduction appropriately, one needs to use singular perturbation methods~\cite{Kuramoto1989, Kuramoto2012, Kunihiro2022}.
The renormalization-group (RG) method in asymptotic analysis provides a general framework for dynamical reduction and for treating singular perturbation problems~\cite{Chen1994PRL, Kunihiro1995PTP, Nozaki2001PRE, Chiba2009, Kunihiro2022}.
In this approach, one first constructs a naive perturbative solution containing secular terms and then absorbs these terms by renormalizing the integration constants. 
The resulting RG flow equations govern the evolution of the slow variables. 
For example, this method has been applied to eliminate momentum from an underdamped Langevin equation with an inhomogeneous temperature~\cite{Matsuo2000PhysicaA} and to derive hydrodynamic equations from the Boltzmann equation~\cite{Kunihiro2006JPhysA}.
Here, we apply the RG method to the coupled equations for the density field $\rho$ and the polarization field $\bm p$ and show that a scalar active field theory can be obtained as an RG flow equation. 
The resulting equation has the same form as that recently discussed in Ref.~\cite{Burekovic2026PRR}.
We also discuss the slow invariant manifold and clarify that the naive adiabatic approximation does not yield the dynamics on this manifold. 

\subsection{Perturbation} 
Our aim is to eliminate the polarization and nematic tensor from Eqs.~\eqref{eq:density_current}, \eqref{eq:polarization}, and \eqref{eq:nematic}.
To eliminate the nematic tensor, one can simply use the quasistationary solution because Eq.~\eqref{eq:nematic} contains only leading-order contributions:
\begin{equation}
Q_{\alpha,\beta} 
= -\frac{\epsilon}{8}\qty[\p_\alpha ( u(\rho)p_\beta )]^\RM{ST} + \frac{\epsilon^2 b_2}{16}\qty[\p_\alpha (\rho\p_\beta\rho)]^\RM{ST} + O(\epsilon^4\nabla^4).
\end{equation}
This solution can also be obtained using perturbation theory.  
Eliminating $Q_{\alpha,\beta}$ from Eqs.~\eqref{eq:density_current} and \eqref{eq:polarization} yields
\begin{equation}
\label{eq:hydrodynamics_ABP}
\begin{split}
\p_{\Frak t}\rho &= - \nabla\cdot \bm J^\rho, 
\\
\bm J^\rho &= \epsilon \bm J^{(1)} 
+ \epsilon^2 \bm J^{(2)} 
+ \epsilon^3 \bm J^{(3)} 
+ \epsilon^4 \bm J^{(4)},
\\
\p_{\Frak t}\bm p &= - \bm p + \epsilon \bm \pi^{(1)} + \epsilon^2 \bm \pi^{(2)} +  \epsilon^3 \bm \pi^{(3)}
\end{split}
\end{equation}
with 
\begin{equation}
\begin{split}
&J_\alpha^{(1)} =   u(\rho)p_\alpha
\\
&J_\alpha^{(2)} = - d(\rho)\p_\alpha\rho 
\\
&J_\alpha^{(3)} = \frac{b_2}{8}\qty[\p_\alpha ( u(\rho)p_\beta )]^\RM{ST} \p_\beta \rho -  c[p_\alpha \nabla^2 
 + 2p_\beta  \p_\beta \p_\alpha]\rho
\\
&J_\alpha^{(4)} = - \frac{b_2^2}{16}\qty[\p_\alpha (\rho\p_\beta\rho)]^\RM{ST}\p_\beta \rho + k \rho \p_\alpha\nabla^2\rho
\end{split}
\end{equation}
and
\begin{align}
\begin{split}
&\pi_\alpha^{(1)} = - h(\rho) \p_\alpha \rho
\\
&\pi_\alpha^{(2)} = \p_\beta \qty{
\eta(\rho) \qty[\p_\alpha p_\beta]^\RM{ST}
+ M_{\alpha,\beta,\mu,\nu}(\rho)p_\mu\p_\nu\rho
}
\\
&\pi_\alpha^{(3)} = \p_\beta\qty{ c\,\rho \qty[\p_\alpha \p_\beta + \frac{1}{2}\delta_{\alpha,\beta}\nabla^2]\rho - \frac{b_2 u(\rho)}{16}  \qty[\p_\alpha(\rho \p_\beta \rho)]^\RM{ST} }.
\end{split}
\end{align}
Here, we have defined
\begin{equation}
\begin{split}
h(\rho) &:= \frac{1-2\rho}{2}, 
\ \ \ \ \ \ \ \ \ \ \ \ 
\eta(\rho) := 2d_0 + \frac{u^2(\rho)}{8},
\\
M_{\alpha,\beta,\mu,\nu}(\rho) &:= 
\qty[\frac{b_2}{4} + \frac{u(\rho)}{16}]\delta_{\alpha,\beta}\delta_{\mu,\nu}
+
\qty[b_0 - \frac{b_2}{4} - \frac{u(\rho)}{16}]\delta_{\alpha,\mu}\delta_{\beta,\nu}
\\
&\ +
\qty[\frac{b_2}{4} - \frac{u(\rho)}{16}]\delta_{\alpha,\nu}\delta_{\beta,\mu}
.
\end{split}
\end{equation}
Note that $\bm J^{(n)} = \bm J^{(n)}[\rho]$ for $n=2,4$ and $\bm \pi^{(n)} = \bm \pi^{(n)}[\rho]$ for $n=1,3$ depend only on $\rho$, whereas $\bm J^{(n)} = \bm J^{(n)}[\rho,\bm p]$ for $n=1,3$ and $\bm \pi^{(2)} = \bm \pi^{(2)}[\rho,\bm p]$ depend on both $\rho$ and $\bm p$. 
The latter quantities are linear in $\bm p$.

We now expand $\rho = \rho(\bm r, \mathfrak t)$ and $\bm p = \bm p(\bm r,\Frak t)$ in powers of $\epsilon$.
To this end, it is convenient to use the fact that Eq.~\eqref{eq:hydrodynamics_ABP} is invariant under $\epsilon \to -\epsilon$ and $\bm p\to  -\bm p$, with $\rho$ unchanged.
The expansions therefore take the form
\begin{align}
\label{eq:perturbation_expansion}
\begin{split}
&\rho = \rho^{(0)}+ \epsilon^2 \rho^{(2)}  +  \epsilon^4 \rho^{(4)} + O(\epsilon^6), \\
& \bm p  = \epsilon  \bm p ^{(1)} +  \epsilon^3  \bm p ^{(3)} + O(\epsilon^5).
\end{split}
\end{align}
In \ref{ape:perturbation}, we calculate the perturbative solution up to $O(\epsilon^4)$ and obtain
\begin{align}
\rho(\bm r, \Frak t; \Frak t_0, \Phi) 
= \Phi 
-\nabla\cdot \qty[
\qty( \epsilon^2 \bm{\mathcal J}^{(2)} [\Phi]
+ \epsilon^4 \bm{\mathcal J}^{(4)}[\Phi])(\Frak t - \Frak t_0 )
+\frac{\epsilon^4}{2} \delta \bm{\mathcal J}^{(2)}[\Phi] \, (\Frak t - \Frak t_0 )^2
 ] + O(\epsilon^6),
 \label{eq:perturbation_solution}
\end{align}
where $\Phi = \Phi(\bm r,\Frak t_0)$ is an integration constant, and each term is defined by
\begin{align}
\bm{\mathcal J}^{(2)}[\Phi] &:= \bm J^{(1)}[\Phi, \bm \pi^{(1)}[\Phi]] + \bm J^{(2)}[\Phi], 
\label{eq:current_2nd_order}
\\
\bm{\mathcal J}^{(4)}[\Phi] &:= \bm J^{(1)}[\Phi, \bm{\Frak p}^{(3)}[\Phi]] + \bm J^{(3)}[\Phi, \bm \pi^{(1)}[\Phi] ]+ \bm J^{(4)}[\Phi],
\label{eq:current_4th_order}
\\
\bm{\Frak p}^{(3)}[\Phi] &:= - \delta \bm \pi^{(1)}[\Phi] + \bm \pi^{(2)}[\Phi, \bm \pi^{(1)}[\Phi]] 
+ \bm \pi^{(3)}[\Phi]. 
\label{eq:non_sec_p3}
\end{align}
For a functional $H[\Phi]$, we have also defined $\delta H[\Phi]$ by
\begin{equation}
\delta H[\Phi] := \int_{V}\dd[2]\bm r'\, F^{(2)}(\bm r')\eval{ \fdv{H[\rho](\bm r)}{\rho(\bm r')} }_{\rho = \Phi}
\end{equation}
with $F^{(2)} = -\nabla\cdot \bm{\mathcal J}^{(2)}$.

\subsection{Renormalization}
\label{subsec:RG}
The solution in Eq.~\eqref{eq:perturbation_solution} contains secular terms and is therefore valid only for $\Frak{t} \sim \Frak{t}_0$. 
A solution valid for $\Frak{t}\sim \Frak t_0 + \Frak{t}'$ can be constructed by writing $\Frak t - \Frak t_0 = (\Frak t - \Frak t_0 - \Frak{t}') + \Frak t'$ and absorbing all terms independent of $(\Frak t - \Frak t_0 - \Frak{t}')$ into the integration constant.
This procedure yields a solution with the new integration constant $\Phi'=\Phi'(\bm r,\Frak t_0,\Frak t')$.
We call this operation ``renormalization'' and denote the transformation from $\Phi(\bm r,\Frak t_0)$ to $\Phi'(\bm r, \Frak t_0, \Frak t')$ by  
\begin{equation}
\hat{\mathcal R}_{\Frak t'}[\Phi] =  \Phi'.
\end{equation}
The transformation is explicitly defined by
\begin{equation}
\hat{\mathcal R}_{\Frak t'}[\Phi] =
\Phi  -\nabla\cdot \qty( \epsilon^2 \bm{\mathcal J}^{(2)} [\Phi]
+ \epsilon^4 \bm{\mathcal J}^{(4)}[\Phi])\, \Frak t' 
 - \frac{\epsilon^4}{2} \nabla \cdot\delta\bm{\mathcal J}^{(2)}[\Phi]\, \Frak t'^2.
 \label{Eq:def_RG}
\end{equation}
This definition implies that the transformation satisfies the semigroup property (see \ref{ape:semigroup} for the derivation):
\begin{equation}
\hat{\mathcal R}_{\Frak t''} \circ \hat{\mathcal R}_{\Frak t'}  = \hat{\mathcal R}_{\Frak t'' + \Frak t'} + O(\epsilon^6).
\label{eq:semigroup}
\end{equation}
We now consider the infinitesimal transformation of $\Phi = \Phi(\bm r, \Frak t)$:
\begin{equation}
\hat{\mathcal R}_{\Delta \Frak t}[\Phi] = \Phi(\bm r,\Frak t) + \Delta \Frak{t} \, \beta[\Phi] + o(\Delta \Frak t),
\end{equation}
where 
\begin{equation}
\beta[\Phi] := \eval{ \dv{\hat{\mathcal R}_{\tau}[\Phi]}{\tau} }_{\tau = 0}.
\end{equation}
Equation~\eqref{eq:semigroup} gives $\hat{\mathcal R}_{\Delta \Frak t}[\Phi]= \Phi'(\bm r, \Frak t, \Delta \Frak t) = \Phi(\bm r, \Frak t + \Delta \Frak t)$.
Taking the limit $\Delta \Frak t\to 0$ yields 
\begin{equation}
\p_{\Frak t} \Phi = \beta[\Phi], \label{eq:flow_equation_formal}
\end{equation}
which is the flow equation for the integration constant. 
From Eq.~\eqref{Eq:def_RG}, we find that this flow equation is the continuity equation up to $O(\epsilon^4)$:
\begin{equation}
\p_{\Frak t} \Phi = - \nabla\cdot \qty(
\epsilon^2 \bm{\mathcal J}^{(2)} + \epsilon^4 \bm{\mathcal J}^{(4)}).
\label{eq:flow_equation}
\end{equation}
Alternatively, requiring $\rho(\bm r, \Frak t; \Frak t_0, \Phi(\bm r, \Frak t_0))$ to be independent of $\Frak t_0$ yields the same equation:
\begin{equation}
\eval{ \dv{\rho(\bm r,\Frak t; \Frak t_0, \Phi(\bm r, \mathfrak t_0))}{\mathfrak t_0} }_{\mathfrak t_0 = \mathfrak t} = 0.
\label{eq:envelope}
\end{equation}
This equation is known as the RG equation or the envelope equation~\cite{Kunihiro2022}.

\subsection{Scalar field theory}
\label{subsec:field_theory_ABP}
We now return to the original slow time variable $t = \epsilon^2 \Frak t$.
The continuity equation becomes $\p_t \Phi = -\nabla \cdot \bm J$ with $\bm J = \bm J(\bm r,t)$ given by
\begin{equation}
\bm J= \bm{\mathcal J}^{(2)} + \epsilon^2 \bm{\mathcal J}^{(4)} + O(\epsilon^4).
\end{equation}
The leading-order current $\bm{\mathcal J}^{(2)}$ is
\begin{equation}
\bm {\mathcal J}^{(2)} = -\mathcal D(\Phi)\nabla \Phi
\end{equation}
with
\begin{equation}
\mathcal D(\Phi) := d(\Phi) + u(\Phi)h(\Phi).
\label{eq:diffusion_coeff_ABP}
\end{equation}
This expression can be written in the form
\begin{equation}
\label{eq:diffusion_free_energy}
 \mathcal D(\Phi)\nabla\Phi  = \nabla\fdv{\mathcal F_0[\Phi]}{\Phi}
\end{equation}
with
\begin{equation}
\mathcal F_0[\Phi] 
= \int_V\dd[2]\bm r\ \qty[\frac{1}{2}\qty(d_0 + \frac{1}{2})\Phi^2 + \frac{1}{6}\qty(b_0 - \frac{3}{2})\Phi^3 + \frac{1}{12}\Phi^4].
\label{eq:bulk_free_energy_ABP}
\end{equation}
Here, $d_0$ is the dimensionless bare diffusion coefficient defined just below Eq.~\eqref{eq:one_body_rescaled}, and $b_0 = \rho_{*}B_0$ is determined by the interaction potential and pair distribution function [see Eq.~\eqref{eq:B0}]. 
Note that the form of the leading-order current $\bm {\mathcal J}^{(2)}$ is the same as that obtained by the adiabatic approximation~\cite{Speck2015JCP}. 

The higher-order current $\bm{\mathcal J}^{(4)}$ can be written as
\begin{equation}
\bm{\mathcal J}^{(4)}= \mathcal K(\Phi) \nabla\nabla^2 \Phi + \mathcal A(\Phi) (\nabla^2 \Phi)(\nabla \Phi) + \mathcal B(\Phi) \nabla \abs{\nabla\Phi}^2 + \mathcal C(\Phi)|\nabla\Phi|^2 \nabla\Phi. 
\label{eq:current_fourth_order}
\end{equation}
Explicit expressions for the coefficients $\mathcal A(\Phi)$, $\mathcal B(\Phi)$, $\mathcal C(\Phi)$, and $\mathcal K(\Phi)$ are given in Appendix~\ref{ape:coefficients_ABP}.
We absorb a higher-order potential term into the effective free energy functional through
\begin{align}
\mathcal F[\Phi] = \mathcal F_0[\Phi] + \frac{\epsilon^2}{2}\int_V\dd[2]\bm r\ \mathcal K(\Phi)\abs{\nabla\Phi}^2
\end{align}
so that the total current $\bm J$ can be written as
\begin{align}
\bm J = - \nabla\fdv{\mathcal F[\Phi]}{\Phi} - \lambda(\Phi) \nabla  |\nabla \Phi|^2 
+ \zeta (\Phi) (\nabla^2 \Phi) \nabla \Phi
+ \nu(\Phi)|\nabla\Phi|^2 \nabla\Phi
\label{eq:current_GAMB+}
\end{align}
with
\begin{align}
\label{eq:nonintegrable_terms}
\begin{split}
&\lambda(\Phi):=\epsilon^2\qty[\frac{1}{2}\mathcal K'(\Phi) - \mathcal B(\Phi)]
,\ \ \ \ \ \ 
\zeta(\Phi) := \epsilon^2\qty[\mathcal A(\Phi) - \mathcal K'(\Phi)]
, \\
&
\nu(\Phi) := \epsilon^2\qty[\mathcal C(\Phi) - \frac{1}{2} \mathcal K''(\Phi)].
\end{split}
\end{align}
Since the deterministic current alone does not uniquely determine the mobility in Eq.~\eqref{eq:current_GAMB+}, we adopt a unit-mobility representation~\cite{Burekovic2026PRR}.
In the original physical units, all the coefficients are independent of $\epsilon$, as shown in \ref{ape:physical_units}.
The $\nu$-term in Eq.~\eqref{eq:current_GAMB+} is absent from AMB+.
We also note that $\Phi$ is not restricted to values near the critical density, and thus the equation $\p_t \Phi = - \nabla\cdot \bm J$ with this current gives the general form of the scalar field theory up to fourth order in spatial gradients, $\nabla^4$.
An equation of the same form has also been derived for quorum-sensing particles, and its general properties have been studied in Ref.~\cite{Burekovic2026PRR}.
Thus, this result implies that ABPs with steric interactions should also share the properties discussed in Ref.~\cite{Burekovic2026PRR}.

We finally remark on the connection to AMB+.
AMB+ retains terms only up to $O(\nabla^3\Phi^2)$, and thus Eq.~\eqref{eq:current_GAMB+} reduces to AMB+ when $\Phi$ is close to a reference density $\varphi_0$.
Indeed, setting 
\begin{equation}
\Phi (\bm r,t)= \varphi_0 + \varphi(\bm r,t)
\end{equation}
and neglecting terms of $O(\nabla^3\varphi^3)$, we find that Eq.~\eqref{eq:current_GAMB+} reduces to
\begin{align}
\bm J = - \nabla\fdv{\mathcal F[\varphi]}{\varphi} - \lambda \nabla |\nabla \varphi|^2 
+ \zeta (\nabla^2 \varphi) \nabla \varphi,
\end{align}
where $\lambda = \lambda(\varphi_0)$ and $\zeta = \zeta(\varphi_0)$ are constants.
The stiffness depends only linearly on $\varphi$: $\kappa = \mathcal K(\varphi_0)+ \mathcal K'(\varphi_0)\varphi$.
The reference density $\varphi_0$ is usually chosen such that the cubic term in the effective bulk free energy, Eq.~\eqref{eq:bulk_free_energy_ABP}, vanishes. 
Then, the coefficient of the quadratic term can become negative and Eq.~\eqref{eq:bulk_free_energy_ABP} can take a double-well form depending on the parameters, leading to motility-induced phase separation.

\subsection{Invariant manifold}
\label{subsec:invariant}
\begin{figure}[t]
\centering
\includegraphics[width=1.0\linewidth]{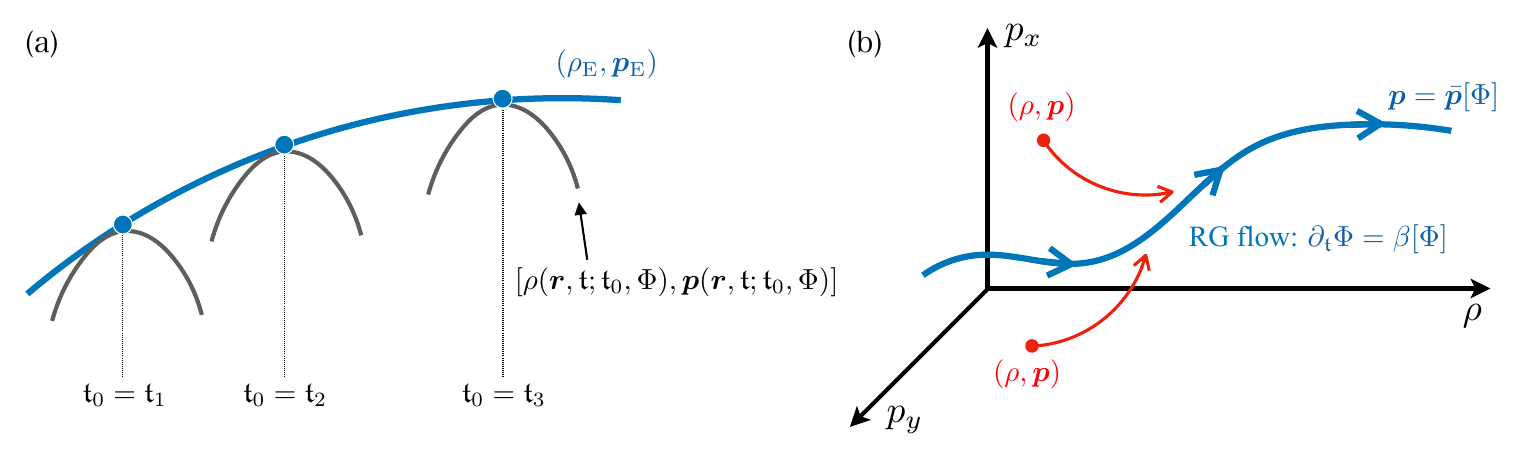}
\caption{\label{fig:RG}(a)~Schematic illustration of the construction of the envelope (blue line) from the local solutions (gray lines) obtained using perturbation theory.
(b)~Schematic of the slow invariant manifold in $(\rho,\bm p)$-space. 
The blue line represents the invariant manifold $\bm p = \bar{\bm p}[\Phi]$, which is parameterized by the density $\Phi$. 
Fast relaxation is schematically depicted by the red arrows. 
The slow dynamics on the invariant manifold is governed by the RG equation.
}
\end{figure}
When the system is described by $(\rho, \bm p)$, it relaxes and its long-time behavior is confined to a subspace.
Such a subspace is called an (attracting) invariant manifold, since a trajectory starting on the manifold remains on it. 
In the current case, the invariant manifold is simply given by $\mathbb M = \{(\rho,\bm p) = (\Phi, \bar{\bm p}[\Phi])\}$.
The RG method also provides a systematic way to read off invariant manifolds~\cite{Ei2000}.
Here, we briefly discuss the invariant manifold for the scalar active field theory.
We also comment on the failure of the adiabatic approximation, which is widely used to eliminate the polarization in scalar active systems.

We first recapitulate the interpretation of the RG equation, Eq.~\eqref{eq:envelope}, as an equation for constructing the envelope~\cite{Kunihiro1995PTP}.
As mentioned in Section~\ref{subsec:RG}, the approximate solution in the long-time limit obtained through naive perturbation theory should be valid only for $\Frak t\sim \Frak t_0$.
Up to $O(\epsilon^4)$, it is given by Eq.~\eqref{eq:perturbation_solution} for the density and 
\begin{align}
\bm p(\bm r,\Frak t; \Frak t_0,\Phi)
= \epsilon\bm \pi^{(1)}[\Phi] + \epsilon^3\qty[ \delta \bm \pi^{(1)}[\Phi](\Frak t-\Frak t_0) + \bm{\Frak p}^{(3)}[\Phi] ]
+O(\epsilon^5)
\end{align}
for the polarization.
Here, $\bm{\Frak p}^{(3)}[\Phi]$ is the non-secular part of $\bm p^{(3)}$ defined by Eq.~\eqref{eq:non_sec_p3}.
We denote the local solution by $\bm P_\RM{local}(\bm r,\Frak t;\Frak t_0, \Phi) = [\rho(\bm r,\Frak t;\Frak t_0,\Phi), \bm p(\bm r,\Frak t; \Frak t_0,\Phi)]$, which is schematically depicted in Figure~\ref{fig:RG}(a).
The RG transformation discussed in Section~\ref{subsec:RG} generates another local solution at a different ``initial'' time $\Frak t_0$ with a new integration constant.
The RG flow equation given by Eq.~\eqref{eq:flow_equation_formal} describes how the integration constant $\Phi$ changes under the RG transformation. 
Using the solution of the flow equation, $\Phi = \Phi(\Frak t)$, one can obtain the envelope of the family of local solutions as
\begin{align}
\bm P_\RM{E}(\bm r, \Frak t) = \bm P_\RM{local}(\bm r,\Frak t;\Frak t, \Phi(\Frak t)).
\end{align}
The envelope $\bm P_\RM{E} = (\rho_\RM{E}, \bm p_\RM{E})$ is schematically shown in Figure~\ref{fig:RG}(a). 

This interpretation of the solution as an envelope allows us to identify $\bar {\bm p}[\Phi]$, which defines the invariant manifold $\mathbb M$~\cite{Ei2000}.
Since the envelope $\bm P_\RM{E}(\bm r,\Frak t)$ satisfies the original dynamical equations, Eq.~\eqref{eq:hydrodynamics_ABP}, up to $O(\epsilon^4)$, $\bar {\bm p}[\Phi]$ can be read off as
\begin{align}
\bar {\bm p}[\Phi] &=
 \epsilon\bm \pi^{(1)}[\Phi] + \epsilon^3 \bm{\Frak p}^{(3)}[\Phi] 
+O(\epsilon^5) \notag \\
& = 
 \epsilon\bm \pi^{(1)}[\Phi] + \epsilon^3 \qty( - \delta \bm \pi^{(1)}[\Phi] + \bm \pi^{(2)}[\Phi, \bm \pi^{(1)}[\Phi]] 
+ \bm \pi^{(3)}[\Phi])
+O(\epsilon^5),
\label{eq:p_bar}
\end{align}
which is simply the non-secular part of the perturbative solution. 
More rigorously, one needs to verify that $\bar {\bm p}[\Phi]$ satisfies the invariance condition~\cite{Gorban2013}. 
We now define $\bm q[\rho,\bm p] = \bm p - \bar{\bm p}[\rho]$.
A direct calculation shows that $\p_{\Frak t}\bm q[\Phi,\bar{\bm p}] = \bm 0 + O(\epsilon^5)$, i.e.,
\begin{align}
\eval{\p_{\Frak t}\bm p}_{\rho=\Phi, \, \bm p=\bar{\bm p}[\Phi]} = \eval{\p_{\Frak t}\bar{\bm p}[\rho]}_{\rho=\Phi, \, \bm p=\bar{\bm p}[\Phi]}
=\int_{V}\dd[2]\bm r'
\eval{\fdv{\bar{\bm p}[\rho](\bm r)}{\rho(\bm r')} }_{\rho=\Phi}\p_{\Frak t}\Phi (\bm r'),
\label{eq:invariance}
\end{align}
where we omit $O(\epsilon^5)$ (see \ref{ape:invariance} for the derivation).
The invariant manifold is schematically depicted in Figure~\ref{fig:RG}(b).
In the long-time limit, the dynamics is confined to the manifold defined by $\bm p = \bar{\bm p}[\Phi]$, and the slow evolution on this manifold is governed by the RG flow equation, which is precisely the scalar field theory.

The invariant-manifold perspective also clarifies why the naive adiabatic approximation yields an incorrect expression for the higher-order current.
In the adiabatic approximation, we simply set $\p_{\Frak t}\bm p=0$ to eliminate the polarization from Eq.~\eqref{eq:hydrodynamics_ABP}. 
At the lowest order, we obtain
\begin{align}
\bm p^\RM{ad}[\rho]\simeq \epsilon \bm \pi^{(1)}[\rho].
\end{align}
Higher-order corrections can be obtained by iteratively substituting this solution into Eq.~\eqref{eq:hydrodynamics_ABP} with $\p_{\Frak t}\bm p=0$ and neglecting terms of $O(\epsilon^5)$:
\begin{align}
\bm p^\RM{ad}[\rho]\simeq \epsilon \bm \pi^{(1)}[\rho] + \epsilon^3 \qty( \bm \pi^{(2)}[\rho, \bm \pi^{(1)}[\rho]] 
+ \bm \pi^{(3)}[\rho]).
\label{eq:adiabatic}
\end{align}
Although substituting this quasi-stationary solution into the density current in Eq.~\eqref{eq:hydrodynamics_ABP} also yields an $O(\nabla^4)$ theory, it does not describe the dynamics on the slow invariant manifold: comparison with Eq.~\eqref{eq:p_bar} shows that $\bm p^\RM{ad}[\rho]$ lacks the term $-\delta\bm \pi^{(1)}[\Phi]$ and thus does not satisfy the invariance condition. 
This highlights the necessity of using a singular perturbation method. 
A recent study~\cite{Burekovic2026PRR} applied multiple timescale analysis, which is also a singular perturbation method, to hydrodynamic equations for a quorum-sensing model and also demonstrated the failure of the adiabatic approximation. 
However, the adiabatic approximation remains useful when we are interested only in the leading order, which corresponds to considering the effective bulk free energy alone. 
A singular perturbation method is needed at higher orders, where terms that break time-reversal symmetry arise.

\section{Scalar field theory for chiral active Brownian particles}
\label{sec:chiral}
In this section, we discuss a scalar field theory for chiral active Brownian particles (cABPs) derived using the method presented above for ABPs.
We show that chirality introduces two additional terms into the $O(\nabla^4)$ field theory derived in the previous section.
Since the derivation follows the same procedure as for ABPs, we provide only an outline. 

\subsection{Setup}
We consider interacting cABPs, whose translational motion is governed by Eq.~\eqref{eq:ABP_posi}.
The orientational dynamics is assumed to obey~\cite{Teeffelen2008PRE, Ma2017SoftMat}
\begin{equation}
\dv{\phi_j(t)}{t} = \Omega + \sqrt{\frac{2}{\tau_\RM{p}}}\eta_j(t), 
\label{eq:cABP_ori}
\end{equation}
where $\eta_j(t)$ is a Gaussian white noise with zero mean and unit variance, $\tau_\RM{p}$ is the persistence time, and $\Omega$ represents the constant chiral torque, or intrinsic angular speed. 
When $\Omega$ is set to zero, the model reduces to the ABP model. 
For finite $\Omega$, each particle undergoes self-propelled motion biased to the left ($\Omega > 0$) or right ($\Omega<0$).

The dynamics of the one-body distribution function for cABPs is given by
\begin{equation}
\p_{\Frak t} \Psi(\bm r,\phi, \Frak t) =  -\nabla\cdot \qty[
 \bm {\mathcal G}(\bm r,\phi,\Frak t)   + \epsilon\bm e(\phi) -  \epsilon^2 d_0\nabla] \Psi(\bm r,\phi,\Frak t) 
-\p_\phi \qty[\omega - \p_\phi]\Psi(\bm r,\phi, \Frak t),
\label{eq:one-body_cABP}
\end{equation}
with
\begin{align}
\omega := \Omega \tau_\RM{p}.
\end{align}
Here, we have rescaled time and length and introduced the small parameter $\epsilon \ll 1$, as described in Section~\ref{subsec:scaling}.
We have also introduced the time variable $\Frak t$ defined in Eq.~\eqref{eq:fast_time}.
We make the same assumptions about the pair distribution function as in Section~\ref{subsec:closure} and use the approximate expression for the conditional force $\bm{\mathcal G}$ given by Eq.~\eqref{eq:con_force_expansion}.
We then obtain a closed equation for the one-body distribution function with the chirality parameter $\omega$.
Note that Eq.~\eqref{eq:con_force_expansion} is applicable only for small $\omega$. 
When $\omega$ is large, the scalar coefficients in Eq.~\eqref{eq:con_force_expansion} should be replaced with tensors (see \ref{ape:gr}).
Here, we focus only on the case of small $\omega$, but the field theory should retain the same form even for large $\omega$.  

\subsection{Effective hydrodynamic equations}
As in the case of ABPs discussed in Section~\ref{sec:hydrodynamics}, one can obtain effective hydrodynamic equations from Eq.~\eqref{eq:one-body_cABP} for the density field, polarization, and nematic tensor. 
Orientational moments of higher rank than the nematic tensor are not relevant for deriving $O(\nabla^4)$ field theories. 
Since the equation for the nematic tensor contains only leading-order terms, one can apply the naive adiabatic approximation, $\p_\Frak{t} \mathsf Q = 0$, to eliminate $\mathsf Q$ (see Section~\ref{sec:hydrodynamics}): 
\begin{align}
Q_{\alpha,\beta} 
= -\frac{\epsilon}{2} (\mathsf W_2^{-1})_{\alpha,\mu}\, \qty[\p_\mu u(\rho)p_\beta]^\RM{ST} + \frac{\epsilon^2 b_2}{4} (\mathsf W_2^{-1})_{\alpha,\mu}\, \qty[\p_\mu (\rho\p_\beta\rho)]^\RM{ST} + O(\epsilon^4\nabla^4),
\label{eq_nematic_cABP}
\end{align}
where we introduce the matrix
\begin{align}
\mathsf W_n = \mqty(
n^2 & n\omega \\ -n\omega & n^2
).
\end{align}
By eliminating the nematic tensor using Eq.~\eqref{eq_nematic_cABP}, we obtain the equations for the density and polarization:
\begin{equation}
\label{eq:hydrodynamics_cABP}
\begin{split}
\p_{\Frak t}\rho &= - \nabla\cdot \bm J^\rho, 
\\
\bm J^\rho &= \epsilon \bm J^{(1)} 
+ \epsilon^2 \bm J^{(2)} 
+ \epsilon^3 \bm J^{(3)} 
+ \epsilon^4 \bm J^{(4)},
\\
\p_{\Frak t}\bm p &= - \mathsf W_1\, \bm p 
+ \epsilon \bm \pi^{(1)} 
+ \epsilon^2 \bm \pi^{(2)} 
+ \epsilon^3 \bm \pi^{(3)},
\end{split}
\end{equation}
with 
\begin{equation}
\begin{split}
&J_\alpha^{(1)} =   u(\rho)p_\alpha
\\
&J_\alpha^{(2)} = - d(\rho)\p_\alpha\rho 
\\
&J_\alpha^{(3)}
= \frac{b_2}{2}(\mathsf W_2^{-1})_{\alpha,\mu}
\qty[\p_\mu\qty(u(\rho)p_\beta)]^\RM{ST}\p_\beta\rho
-c[p_\alpha \nabla^2 
 + 2p_\beta  \p_\beta \p_\alpha]\rho
\\
&J_\alpha^{(4)}
= -\frac{b_2^2}{4}(\mathsf W_2^{-1})_{\alpha,\mu}
\qty[\p_\mu\qty(\rho\p_\beta\rho)]^\RM{ST}\p_\beta\rho
+k\rho\p_\alpha\nabla^2\rho
\end{split}
\end{equation}
and
\begin{equation}
\begin{split}
&\pi_\alpha^{(1)} = - h(\rho) \p_\alpha \rho
\\
&\pi_\alpha^{(2)} 
= 
\p_\beta\qty{
\eta_{\alpha,\mu}(\rho)\qty[\p_\mu p_\beta]^\RM{ST}
+ M_{\alpha,\beta,\mu,\nu}(\rho)p_\mu\p_\nu\rho
}
\\
&\pi_\alpha^{(3)} = \p_\beta\qty{
c\,\rho \qty[\p_\alpha \p_\beta + \frac{1}{2}\delta_{\alpha,\beta}\nabla^2]\rho - \frac{b_2u(\rho)}{4}(\mathsf W_2^{-1})_{\alpha,\mu}\qty[\p_\mu\qty(\rho\p_\beta\rho)]^\RM{ST}
},
\end{split}
\end{equation}
where we define
\begin{equation}
\begin{split}
h(\rho) &:= \frac{1-2\rho}{2},
\\
\eta_{\alpha,\beta}(\rho)
&
:=\qty[2d_0+\frac{u^2(\rho)}{8+2\omega^2}]\delta_{\alpha,\beta}
-\frac{\omega u^2(\rho)}{16+4\omega^2}\varepsilon_{\alpha,\beta},
\\
M_{\alpha,\beta,\mu,\nu}(\rho)
&:= \qty[\frac{b_2}{4}+\frac{u(\rho)}{16+4\omega^2}]
\delta_{\alpha,\beta}\delta_{\mu,\nu} + \qty[b_0-\frac{b_2}{4}-\frac{u(\rho)}{16+4\omega^2}]
\delta_{\alpha,\mu}\delta_{\beta,\nu}
\\
&\ 
+\qty[\frac{b_2}{4}-\frac{u(\rho)}{16+4\omega^2}]
\delta_{\alpha,\nu}\delta_{\beta,\mu}
+\frac{\omega u(\rho)}{32+8\omega^2}\qty(
\varepsilon_{\alpha,\mu}\delta_{\beta,\nu}
+\varepsilon_{\alpha,\nu}\delta_{\beta,\mu}
-\varepsilon_{\alpha,\beta}\delta_{\mu,\nu}
).
\label{eq:coefficients_hydro_cABP}
\end{split}
\end{equation}
Here, $\varepsilon_{\alpha,\beta}$ denotes the Levi--Civita tensor in two dimensions whose components are given by $\varepsilon_{x,x} = \varepsilon_{y,y} = 0$ and  $\varepsilon_{x,y} = - \varepsilon_{y,x} = 1$.
When $\omega =0$, these equations reduce to the effective hydrodynamic equations for ABPs given by Eq.~\eqref{eq:hydrodynamics_ABP}.

\subsection{Chiral active field theory}
\label{subsec:field_theory_cABP}
As discussed in Section~\ref{sec:RG}, one can eliminate the fast variable, i.e., the polarization, from the effective hydrodynamic equations using a singular perturbation method. 
We expand $\rho$ and $\bm p$ as in Eq.~\eqref{eq:perturbation_expansion}.
Since Eq.~\eqref{eq:hydrodynamics_cABP} has the same form as Eq.~\eqref{eq:hydrodynamics_ABP}, except that the coefficient of the damping term in $\bm p$ is the matrix $\mathsf W_1$, whose antisymmetric part produces oscillations, the formal perturbative solution up to $O(\epsilon^4)$ is given by Eq.~\eqref{eq:perturbation_solution}.
The currents $\bm{\mathcal J}^{(2)}$ and $\bm{\mathcal J}^{(4)}$ retain the forms of Eqs.~\eqref{eq:current_2nd_order} and \eqref{eq:current_4th_order}, respectively, with the replacement $\bm\pi^{(n)} \to \mathsf W_1^{-1}\bm\pi^{(n)}$, except that $\delta\bm\pi^{(1)}$ in Eq.~\eqref{eq:non_sec_p3} is replaced by $\mathsf W_1^{-2}\delta\bm\pi^{(1)}$.
We then apply the RG method and obtain the flow equation for the integration constant in the same way as in Section~\ref{subsec:RG}.
The effects of chirality appear in the explicit forms of the currents in Eq.~\eqref{eq:flow_equation}.
The leading-order current $\bm {\mathcal J}^{(2)}$ is given by
\begin{align}
\bm {\mathcal J}^{(2)} = - \mathcal D(\Phi)\nabla\Phi - \mathcal D_\perp(\Phi)\nabla^\perp\Phi,
\end{align}
where $\nabla^\perp = (\p_y,-\p_x)$ and the diffusion coefficients are defined by
\begin{align}
\mathcal D(\Phi) = d(\Phi) + \frac{u(\Phi)h(\Phi)}{1+\omega^2}
,\ \ \ \ \ 
\mathcal D_\perp(\Phi) = -\frac{\omega u(\Phi)h(\Phi)}{1+\omega^2}.
\end{align}
The first term represents normal diffusion and can be written in terms of an effective bulk free energy, as in Eq.~\eqref{eq:diffusion_free_energy}.
The second term involving $\mathcal D_\perp(\Phi)$ represents ``odd" diffusion.
The odd diffusion term is divergence-free, i.e., $\nabla\cdot [\mathcal D_\perp(\Phi)\nabla^\perp\Phi]=0$, and thus does not affect the density profile, but it generates currents perpendicular to the density gradient~\cite{Hargus2021PRL, Fruchart2023AnnuRev}.
The higher-order current $\bm {\mathcal J}^{(4)}$ can be written as
\begin{align}
  \bm {\mathcal J}^{(4)}
= \mathsf K(\Phi) \nabla\nabla^2 \Phi 
 + \mathsf A(\Phi) (\nabla^2 \Phi)(\nabla \Phi) 
 + \mathsf  B(\Phi) \nabla \abs{\nabla\Phi}^2
 +\mathsf C(\Phi) \abs{\nabla\Phi}^2(\nabla \Phi),
\end{align}
where $\mathsf K(\Phi)$, $\mathsf A(\Phi)$, $\mathsf B(\Phi)$, and $\mathsf C(\Phi)$ are matrices that have the following form:
\begin{align}
X_{\alpha,\beta}(\Phi) = \mathcal X(\Phi)\delta_{\alpha,\beta} + \mathcal X_\perp(\Phi)\varepsilon_{\alpha,\beta}.
\end{align}
The explicit expressions for each element are given in Appendix~\ref{ape:coefficients_cABP}. 
The total current $\bm J = \bm {\mathcal J}^{(2)} + \epsilon^2 \bm {\mathcal J}^{(4)}$ can be decomposed into parallel and perpendicular parts:
\begin{align}
\bm J = \bm J^\parallel + \bm J^\perp.
\end{align}
The parallel part reads
\begin{equation}
\bm J^\parallel = - \nabla\fdv{\mathcal F[\Phi]}{\Phi} - \lambda (\Phi) \nabla  |\nabla \Phi|^2 
+ \zeta  (\Phi) (\nabla^2 \Phi) \nabla \Phi
+ \nu (\Phi)|\nabla\Phi|^2 \nabla\Phi,
\end{equation}
where the effective free-energy functional is given by
\begin{align}
\mathcal F[\Phi] = \mathcal F_0[\Phi] + \frac{\epsilon^2}{2}\int_V\dd[2]\bm r\ \mathcal K(\Phi)\abs{\nabla\Phi}^2,
\end{align}
with
\begin{equation}
\mathcal F_0[\Phi] 
= \int_V\dd[2]\bm r \qty[
\frac{1}{2} \qty{ d_0 + \frac{1}{2(1+\omega^2)} }\Phi^2
+
\frac{1}{6} \qty{b_0-\frac{3}{2(1+\omega^2)}}\Phi^3
+
\frac{1}{12(1+\omega^2)}\Phi^4
].
\end{equation}
The coefficients of the nonintegrable terms, $\lambda(\Phi)$, $\zeta(\Phi)$, and $\nu(\Phi)$, are defined by Eq.~\eqref{eq:nonintegrable_terms}.
The perpendicular current contains several divergence-free terms, including the odd diffusion term, that do not contribute to the dynamics of the scalar field.
We thus find that the field $\Phi$ obeys 
\begin{align}
\p_t \Phi = -\nabla\cdot \bm J^\parallel - \zeta_\perp(\Phi)\qty{\nabla^2\Phi, \Phi} - \nu_\perp(\Phi)\qty{|\nabla\Phi|^2,\Phi},
\label{eq:field_theory_cABP}
\end{align}
where
\begin{align}
\zeta_\perp(\Phi) := \epsilon^2\qty[\mathcal A_\perp(\Phi) - \mathcal K'_\perp(\Phi)]
,\ \ \ \ \ 
\nu_\perp(\Phi) := \epsilon^2[ \mathcal C_\perp(\Phi) - \mathcal B'_\perp(\Phi) ],
\end{align}
and 
\begin{align}
\qty{A,B} := (\p_x A) (\p_y B) - (\p_y A) (\p_xB).
\end{align}
In the limit of vanishing $\omega$, the chiral contributions $\zeta_\perp$ and $\nu_\perp$ vanish, and Eq.~\eqref{eq:field_theory_cABP} reduces to the ABP field theory in Section~\ref{subsec:field_theory_ABP}.
The $\zeta_\perp$- and $\nu_\perp$-terms stem from chirality and can affect the density profiles.
Studying the effects of these two additional terms is left for future work.
Note that even the parallel part $\mathcal X(\Phi)$ depends on the chirality $\omega$ (see Appendix~\ref{ape:coefficients_cABP}), i.e., $\omega$ also affects the parallel current $\bm J^\parallel$. 

\section{Conclusions}
\label{sec:conclusions}
In this paper, we have revisited the microscopic derivation of scalar active field theories for active Brownian particles (ABPs) interacting via a two-body potential in two dimensions. 
We started from the standard approach to constructing hydrodynamic equations for ABPs, namely, applying closure approximations to the equation for the one-body distribution function~\cite{Bialk2013EPL, Speck2015JCP}. 
We then employed a pressure expansion to express the conditional force in terms of the orientational moments and spatial gradients.
By introducing a small parameter for scale separation, we identified which terms must be retained to derive a scalar field theory up to fourth order in spatial gradients. 
This analysis led to coupled equations for the density field and polarization that contain all the terms required for an $O(\nabla^4)$ theory within the closure approximation employed here. 
To eliminate the polarization and obtain an equation for the density field alone, we used the renormalization group (RG) method in the context of singular perturbation theory. 
We have shown that a scalar field theory can be obtained as an RG flow equation. 
The resulting equation has the same form as that proposed in Ref.~\cite{Burekovic2026PRR} for quorum-sensing active particles. 
We have also studied chiral ABPs and shown that the resulting $O(\nabla^4)$ theory contains two additional ``odd'' terms. 
Since our derivation is fully systematic once the closure approximation is adopted, the framework developed here could also be useful for other types of active matter systems. 

Several issues remain to be addressed. 
One concern is the validity of the closure approximation.
We assumed several properties of the pair distribution function to express the interaction integral in terms of the one-body distribution function.
The crudest assumption we made is the quasi-stationary approximation for the pair distribution function.
The conditions under which this assumption is valid should be tested in future work. 
Theoretically, it may be possible to improve the closure approximation by decomposing the interaction integral into adiabatic and superadiabatic parts~\cite{Schmidt2022RMP}.
Another issue is the quantitative comparison of the coefficients in the scalar field theories with data from particle simulations. 
Although we have obtained expressions for all the coefficients in terms of microscopic parameters, their correctness cannot be guaranteed without independent reference values for comparison.
Data-driven parameter estimation would be an important direction for future work to address this issue~\cite{Colen2021PNAS, Joshi2022PRL, Supekar2023PNAS, Hem2025SoftMat}. 
Furthermore, the resulting expressions for the coefficients are complicated and do not offer a clear physical interpretation of each term. 
Further theoretical studies are needed to understand scalar active field theories from a microscopic perspective.

\ack
The authors acknowledge support by the state of Baden-Württemberg through bwHPC and the German Research Foundation (DFG) through grant INST 35/1597-1 FUGG.
Y.K. acknowledges support from the Japan Society for the Promotion of Science (JSPS) through the Overseas Research Fellowships.

\appendix
\renewcommand{\thesubsection}{\Alph{section}.\arabic{subsection}}
\addtocontents{toc}{\protect\appendixtocformat}
\section{Pair distribution function}
\label{ape:gr}
In this appendix, we numerically compute the pair distribution function to investigate the applicability of the assumption given by Eq.~\eqref{eq:assum_gr_3}. 
The setup of the numerical simulations is as follows.
We simulate interacting (chiral) ABPs obeying Eqs.~\eqref{eq:ABP_posi} and \eqref{eq:cABP_ori}. 
For the pairwise interaction, we employ the Weeks--Chandler--Andersen (WCA) potential given by
\begin{align}
U(r) = 4U_0\left[ \left( \frac{\sigma}{r }\right)^{12} - \left( \frac{\sigma}{r }\right)^{6} + \frac{1}{4} \right]
\end{align}
for $r < 2^{1/6}\sigma$ and $U(r)=0$ otherwise, where $r_\RM{cut} = 2^{1/6}\sigma$ is the cutoff length and $\sigma$ denotes the diameter of a particle. 
We choose $\tau = \sigma^2/D_0$ and $\sigma$ as the units of time and length, respectively. 
The control parameters are $v_0\sigma/D_0$, $\tau_\RM{p}D_0/\sigma^2$, $\Omega\sigma^2/D_0$, $\mu U_0/D_0$, and the packing fraction $\varphi_\RM{pack}=\pi\sigma^2 N/(4L^2)$, where $L$ is the side length of the simulation box.  
In the following, we express all quantities in the units defined above.
Here, we examine the effect of chirality on the pair distribution function. 
We focus on states at a relatively low density, $\varphi_\RM{pack} = 0.2$, and fix the parameters at $v_0=50$, $\tau_\RM{p}=1/3$, and $\mu U_0/D_0 = 1\times10^2$.
The total number of particles is $N=5\times 10^3$.
We impose periodic boundary conditions and numerically integrate Eqs.~\eqref{eq:ABP_posi} and \eqref{eq:cABP_ori} using the Euler--Maruyama method with a time step of $\Delta t = 5 \times 10^{-6}$.

\begin{figure}[t]
\centering
\includegraphics[width=1.0\linewidth]{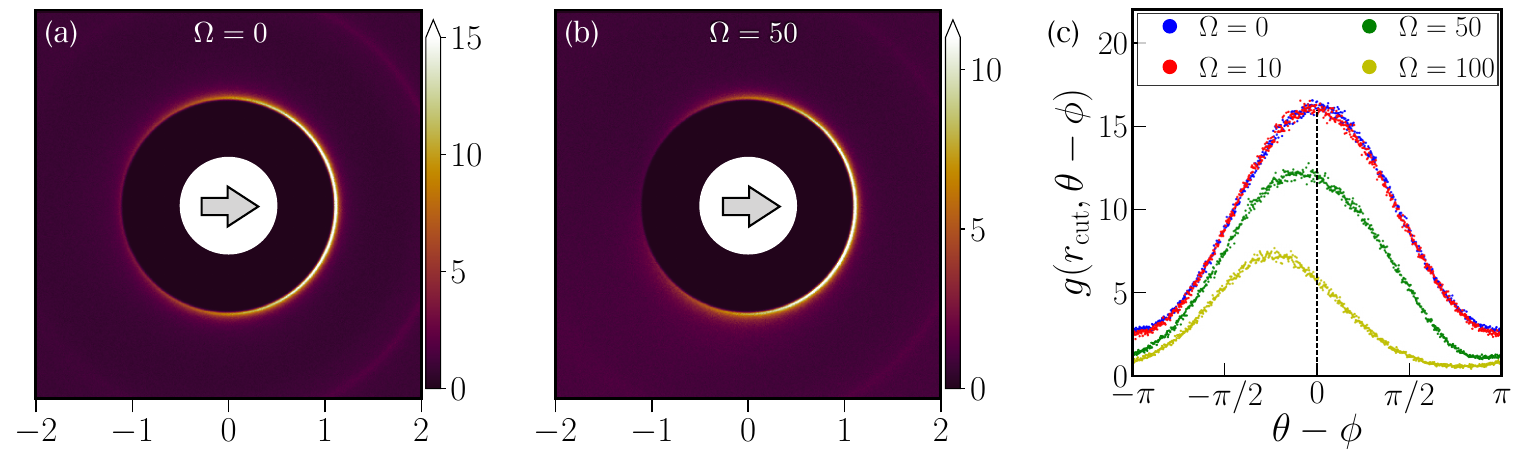}
\caption{\label{fig:gr}The pair distribution function $g(\bm r,\phi)$ in homogeneous states at (a)~$\Omega = 0$ and (b)~$\Omega=50$.
We have measured it in a coordinate system attached to each tagged particle, with the $x$-axis parallel to the orientation $\bm e(\phi)$, indicated by the gray arrow. 
The white circle represents a tagged particle of diameter $\sigma$.
(c)~$g(r,\theta - \phi)$ at $r=r_\RM{cut} = 2^{1/6}\sigma$ for several values of $\Omega$.
Here, $\theta$ denotes the angle of $\bm r$ with respect to the $x$-axis.
The other parameters are set to $\varphi_\RM{pack} = 0.2$, $v_0=50$, and $\tau_\RM{p} = 1/3$.
}
\end{figure}

Figures~\ref{fig:gr}(a) and (b) show the pair distribution function calculated in homogeneous states at $\Omega = 0$ and $\Omega = 50$, respectively.
When $\Omega = 0$, the system consists of ABPs, and $g(\bm r,\phi)$ is symmetric about the axis parallel to the orientation $\bm e(\phi)$ and takes higher values in front of the tagged particle. 
This means that other particles are more likely to be found in front of the tagged particle. 
When $\Omega$ is large, however, $g(\bm r,\phi)$ is no longer symmetric about the axis parallel to the orientation $\bm e(\phi)$, but its axis of symmetry is rotated.
Intuitively, this stems from the fact that particles are less likely to be inside the circular trajectory of the tagged particle.
In Figure~\ref{fig:gr}(c), we also show $g(r,\theta-\phi)$ at the cutoff length $r=r_\RM{cut}$ as a function of $\theta-\phi$ for several values of $\Omega$.
The peak is located at $\theta = \phi$ when $\Omega = 0$ but shifts as $\Omega$ increases.

Given these observations, assuming Eq.~\eqref{eq:assum_gr_3} is reasonable for ABPs ($\Omega = 0$) and cABPs with small $\Omega$.
However, when $\Omega$ is large, Eq.~\eqref{eq:assum_gr_3} should be modified to
\begin{equation}
g(\bm r,\phi) = \tilde g(r,\theta - \phi - \theta_{*}),
\end{equation}
where $\theta_*$ depends on $\Omega$.
This modification replaces the scalar coefficients in the expansion of the conditional force with tensors.
For example, $\gamma \bm e(\phi)$ in Eq.~\eqref{eq:con_force_expansion} becomes 
\begin{equation}
\gamma e_{\alpha}(\phi + \theta_*) = 
\qty(\gamma_\parallel\delta_{\alpha,\beta} + \gamma_\perp\varepsilon_{\alpha,\beta})e_{\beta}(\phi)
\label{eq:assum_gr_4}
\end{equation}
with $\gamma_\parallel = \gamma\cos\theta_*$ and $\gamma_\perp = -\gamma\sin\theta_*$. 
This introduces a perpendicular component into the density-dependent velocity in Eq.~\eqref{eq:effective_speed}~\cite{Kuroda2026}.
Although using Eq.~\eqref{eq:assum_gr_4} alters the pressure expansion in the main text, one could still eliminate the fast variables using the same procedure, and the resulting scalar field theory is expected to have the same form as Eq.~\eqref{eq:field_theory_cABP}. 

\section{Derivation of Equation~\eqref{eq:con_force_expansion}}
\label{ape:pressure_expansion}
In this Appendix, we derive Eq.~\eqref{eq:con_force_expansion} by calculating each coefficient in Eq.~\eqref{eq:gradient_expansion} given by
\begin{align}
&g^{(0)}_\alpha(\phi)
= \frac{\mu \tau_\RM{p}}{\ell_\RM{p}} \int_{r < r_\RM{c}}\dd[2]\bm r\  g(\bm r, \phi)U'(r)\frac{x_\alpha}{r}, \\
& g_{\alpha,\beta}^{(1)}(\phi)
= \frac{\mu \tau_\RM{p}}{\ell_\RM{p}^2} \int_{r < r_\RM{c}}\dd[2]\bm r\  g(\bm r, \phi)U'(r)\frac{x_\alpha x_\beta}{r} ,\\
&g_{\alpha,\beta,\mu}^{(2)}(\phi)
=   \frac{1}{2} \frac{\mu \tau_\RM{p}}{\ell_\RM{p}^3}\int_{r < r_\RM{c}}\dd[2]\bm r\  g(\bm r, \phi)U'(r)\frac{x_\alpha x_\beta x_\mu}{r} , \\
&g_{\alpha,\beta,\mu,\nu}^{(3)}(\phi)
=  \frac{1}{3!} \frac{\mu \tau_\RM{p}}{\ell_\RM{p}^4}\int_{r < r_\RM{c}}\dd[2]\bm r\  g(\bm r, \phi)U'(r)\frac{x_\alpha x_\beta x_\mu x_\nu}{r} .
\end{align}
Here, we have changed the integration variable to $\bm r = \bm s \ell_\RM{macro}$.
In the following, we write $x_\alpha = r e_\alpha(\theta)$.
To calculate each coefficient, we use the assumption expressed by Eq.~\eqref{eq:assum_gr_3}.
It is also convenient to use the integral formula
\begin{equation}
\int_{-\pi}^{\pi}\dd\theta\ f(\theta-\phi)[\bm e^{\otimes n}(\theta)]^\RM{ST} = [\bm e^{\otimes n}(\phi)]^\RM{ST}\int_{-\pi}^{\pi}\dd\theta\ f(\theta)\cos(n\theta),
\label{eq:moment_formura}
\end{equation}
for the function $f(\cdot)$ satisfying $f(\theta) = f(-\theta)$.
This can be easily proven by using 
\begin{equation}
[\underbrace{e_x(\phi)e_x(\phi)\cdots e_x(\phi)}_{n-k}  \underbrace{e_y(\phi) e_y(\phi) \cdots e_y(\phi)}_{k}]^\RM{ST} 
= \frac{1}{2^{n-1}}
 \begin{cases}
    (-1)^{k/2}\cos(n\phi) & \text{$k$: even}, \\
    (-1)^{(k-1)/2}\sin(n\phi) & \text{$k$: odd}.
  \end{cases}
\end{equation}

\vspace{5mm}
\noindent\underline{$g_\alpha^{(0)}(\phi)$:}
One can directly calculate as
\begin{align}
g^{(0)}_\alpha(\phi) &=\frac{\mu \tau_\RM{p}}{\ell_\RM{p}}  \int_{0}^{r_\RM{c}}\dd r\int_{-\pi}^{\pi}\dd\theta\ rU'(r) \tilde g(r,\theta - \phi)e_\alpha(\theta) \notag \\
&= -\gamma e_\alpha(\phi)
\end{align}
with 
\begin{equation}
\gamma := - \frac{\mu \tau_\RM{p}}{\ell_\RM{p}}\int_{0}^{r_\RM{c}}\dd r\int_{-\pi}^{\pi}\dd\theta\ rU'(r) \tilde g(r,\theta)\cos\theta.
\end{equation}

\vspace{5mm}
\noindent \underline{$g^{(1)}_{\alpha,\beta}(\phi)$:}
Using the decomposition
\begin{equation}
e_\alpha(\theta)e_\beta(\theta) = q_{\alpha,\beta}(\theta) + \frac{1}{2}\delta_{\alpha,\beta},
\end{equation}
where $q_{\alpha,\beta}(\theta):= [e_\alpha(\theta)e_\beta(\theta)]^\RM{ST}$, we obtain
\begin{equation}
g_{\alpha,\beta}^{(1)}(\phi) =  - B_0\delta_{\alpha,\beta} - B_2 q_{\alpha,\beta}(\phi),
\end{equation}
with 
\begin{align}
&B_0 
:= - \frac{1}{2} \frac{\mu \tau_\RM{p}}{\ell_\RM{p}^2} \int_{0}^{r_\RM{c}}\dd r\int_{-\pi}^{\pi}\dd\theta\ r^2U'(r) \tilde g(r,\theta), \label{eq:B0}\\
&B_2
:= - \frac{\mu \tau_\RM{p}}{\ell_\RM{p}^2} \int_{0}^{r_\RM{c}}\dd r\int_{-\pi}^{\pi}\dd\theta\ r^2U'(r) \tilde g(r,\theta) \cos(2\theta).
\end{align}

\vspace{5mm}
\noindent\underline{$g^{(2)}_{\alpha,\beta,\mu}(\phi)$:}
We decompose the tensor $e_\alpha(\theta) e_\beta(\theta)e_\mu(\theta)$ as
\begin{equation}
e_\alpha(\theta) e_\beta(\theta)e_\mu(\theta) = 
S_{\alpha,\beta,\mu}^{(3)}(\theta) + [e_\alpha(\theta) e_\beta(\theta)e_\mu(\theta)]^\RM{ST} ,
\end{equation}
where
\begin{align}
S_{\alpha,\beta,\mu}^{(3)}(\theta) := \frac{1}{4}\qty[
e_\alpha(\theta)\delta_{\beta,\mu} + e_\beta(\theta)\delta_{\mu, \alpha} + e_\mu(\theta)\delta_{\alpha,\beta}
].
\end{align}
Using Eq.~\eqref{eq:moment_formura}, we get
\begin{equation}
g_{\alpha,\beta,\mu}^{(2)}(\phi) = -C_1 S_{\alpha,\beta,\mu}^{(3)}(\phi) - C_3 [e_\alpha(\phi) e_\beta(\phi)e_\mu(\phi)]^\RM{ST}
\end{equation}
with
\begin{align}
C_1
&:=
-\frac{1}{2} \frac{\mu \tau_\RM{p}}{\ell_\RM{p}^3}
\int_0^{r_c} \dd r
\int_{-\pi}^{\pi} \dd\theta\ 
r^3 U'(r)\tilde{g}(r,\theta)\cos\theta,
\\
C_3
&:=
-\frac{1}{2} \frac{\mu \tau_\RM{p}}{\ell_\RM{p}^3}
\int_0^{r_c} \dd r
\int_{-\pi}^{\pi} \dd\theta\ 
r^3 U'(r)\tilde{g}(r,\theta)\cos(3\theta).
\end{align}

\vspace{5mm}
\noindent\underline{$g^{(3)}_{\alpha,\beta,\mu,\nu}(\phi)$:}
The tensor $e_\alpha(\theta) e_\beta(\theta)e_\mu(\theta)e_\nu(\theta)$ can be decomposed as
\begin{equation}
e_\alpha(\theta) e_\beta(\theta)e_\mu(\theta)e_\nu(\theta) = \frac{1}{8}\Delta_{\alpha,\beta,\mu,\nu} + S^{(4)}_{\alpha,\beta,\mu,\nu}(\theta) + [e_\alpha(\theta) e_\beta(\theta)e_\mu(\theta)e_\nu(\theta)]^\RM{ST},
\end{equation}
where 
\begin{equation}
\Delta_{\alpha,\beta,\mu,\nu} = \delta_{\alpha,\beta}\delta_{\mu,\nu} + \delta_{\alpha,\mu}\delta_{\beta,\nu} + \delta_{\alpha,\nu}\delta_{\beta,\mu}
\end{equation}
and 
\begin{equation}
\begin{split}
S^{(4)}_{\alpha,\beta,\mu,\nu}(\theta) 
= \frac{1}{6}\Big[
\delta_{\alpha,\beta}q_{\mu,\nu}(\theta) 
+ \delta_{\alpha,\mu}q_{\beta,\nu}(\theta)
+ \delta_{\alpha,\nu}q_{\beta,\mu}(\theta)
+ \delta_{\beta,\mu}q_{\alpha,\nu}(\theta)&
\\
+ \delta_{\beta,\nu}q_{\alpha,\mu}(\theta)
+ \delta_{\mu,\nu}q_{\alpha,\beta}(\theta)&
\Big].
\end{split}
\end{equation}
Using Eq.~\eqref{eq:moment_formura}, we get
\begin{align}
g_{\alpha,\beta,\mu,\nu}^{(3)}(\phi) = - K_0 \Delta_{\alpha,\beta,\mu,\nu} - K_2 S^{(4)}_{\alpha,\beta,\mu,\nu}(\phi) - K_4 [e_\alpha(\phi) e_\beta(\phi)e_\mu(\phi)e_\nu(\phi)]^\RM{ST}
\end{align}
with
\begin{align}
K_0
&:=
-\frac{1}{48}\frac{\mu \tau_\RM{p}}{\ell_\RM{p}^4}
\int_0^{r_c} \dd r
\int_{-\pi}^{\pi} \dd \theta\ 
r^4 U'(r)\tilde{g}(r,\theta), \label{eq:K0}
\\
K_2
&:=
-\frac{1}{6}\frac{\mu \tau_\RM{p}}{\ell_\RM{p}^4}
\int_0^{r_c} \dd r
\int_{-\pi}^{\pi} \mathrm{d}\theta\ 
r^4 U'(r)\tilde{g}(r,\theta)\cos(2\theta),
\\
K_4
&:=
-\frac{1}{6}\frac{\mu \tau_\RM{p}}{\ell_\RM{p}^4}
\int_0^{r_c} \dd r
\int_{-\pi}^{\pi} \dd\theta\ 
r^4 U'(r)\tilde{g}(r,\theta)\cos(4\theta).
\end{align}

Recall that we seek an expression for $\bm{\mathcal G}$ that yields the pressure gradient $\nabla\cdot \mathsf \Pi^\rho$ up to $O(\epsilon^4\nabla^3\rho)$. 
Since $\mathsf Q_n = \int\dd\phi\ [\bm e^{\otimes n}(\phi)]^\RM{ST}\Psi = O(\epsilon^n\nabla^n)$ (see Section~\ref{subsec:order_est}), e.g., $\bm p = O(\epsilon \nabla)$ and $\mathsf Q = O(\epsilon^2\nabla^2)$, the $C_3$-term contributes at $O(\epsilon^6\nabla^5)$, the $K_2$-term at $O(\epsilon^6\nabla^5)$, and the $K_4$-term at $O(\epsilon^8\nabla^7)$.
We thus neglect these three terms and obtain Eq.~\eqref{eq:con_force_expansion} in the main text with $C := C_1/4$ and $K:= -3K_0$.

We remark on the coefficient $K$, which corresponds to a passive contribution to the stiffness.
As is evident from Eq.~\eqref{eq:K0}, $K_0$ is positive when the potential is purely repulsive, and thus $K = - 3K_0$ is negative. 
While counterintuitive, such a negative density-gradient coefficient has also been reported in molecular simulations of a repulsive binary fluid~\cite{Denniston2004PRE}.
On physical grounds, we expect the resulting stiffness in the total current, $\mathcal K(\Phi)$ [Eq.~\eqref{eq:stiffness}], to be sufficient for stable steady states, although a formal proof is beyond the scope of this work.

\section{Perturbative solutions}
\label{ape:perturbation}
In this Appendix, we derive the naive perturbative solution, Eq.~\eqref{eq:perturbation_solution}.
We substitute the expansions of $\rho$ and $\bm p$ into Eq.~\eqref{eq:hydrodynamics_ABP} and determine their coefficients order by order in $\epsilon$ for $\Frak{t} \gg 1$.

\vspace{5mm}
\noindent \underline{$O(\epsilon^0)$:}
At zeroth order, the density equation contains no current and reduces to $\p_{\Frak t} \rho^{(0)} = 0$.
Its solution is
\begin{equation}
\rho^{(0)}(\bm r,\Frak t) = \Phi(\bm r, \Frak t_0).
\end{equation}
The constant of integration $\Phi$ is an arbitrary spatial profile that is independent of $\Frak t$ at this order.
Here, $\Frak t_0$ is the reference or ``initial'' time about which the local perturbative solution is constructed.

\vspace{5mm}
\noindent \underline{$O(\epsilon^1)$:}
At first order, the polarization obeys
\begin{equation}
\p_{\Frak t} \bm p^{(1)} = - \bm p^{(1)} + \bm \pi^{(1)}[\Phi].
\end{equation}
For fixed $\Phi$, the source term $\bm \pi^{(1)}[\Phi]$ on the right-hand side is independent of $\Frak t$.
The general solution also contains a term proportional to $\exp[-(\Frak t-\Frak t_0)]$.
After this fast transient has decayed, the polarization is given by
\begin{equation}
\bm p^{(1)} = \bm \pi^{(1)}[\Phi].
\end{equation}

\vspace{5mm}
\noindent \underline{$O(\epsilon^2)$:}
Substituting the leading-order polarization into the density equation gives
\begin{equation}
\p_{\Frak t} \rho^{(2)} = - \nabla\cdot\qty(
 \bm J^{(1)}[\Phi, \bm \pi^{(1)}] + \bm J^{(2)}[\Phi]
).
\end{equation}
The right-hand side is independent of $\Frak t$, and integration therefore yields
\begin{equation}
 \rho^{(2)} = - (\Frak t - \Frak t_0)\nabla\cdot \bm{\mathcal J}^{(2)} .
\end{equation}
Here, $\bm{\mathcal J}^{(2)}$ is defined as
\begin{align}
\bm{\mathcal J}^{(2)} := \bm J^{(1)}[\Phi, \bm \pi^{(1)}] + \bm J^{(2)}[\Phi].
\end{align}
A time-independent integration field has been absorbed into $\Phi$ in $\rho^{(0)}$.
The term proportional to $\Frak t-\Frak t_0$ is the first secular contribution to the naive perturbative solution.

\vspace{5mm}
\noindent \underline{$O(\epsilon^3)$:}
At third order, the second-order change in the density leads to a variation of the first-order polarization.
Together with the remaining third-order sources, we obtain
\begin{equation}
\p_{\Frak t} \bm p^{(3)} 
= - \bm p^{(3)} + (\Frak t - \Frak t_0)\, \delta \bm \pi^{(1)}[\Phi] 
+\bm \pi^{(2)}[\Phi, \bm \pi^{(1)}] 
+ \bm \pi^{(3)}[\Phi],
\end{equation}
where 
\begin{equation}
\delta \bm \pi^{(1)}[\Phi]  
:=  \int_{V}\dd[2]\bm r'\ F^{(2)}(\bm r')\eval {\fdv{\bm \pi^{(1)}[\rho](\bm r)}{\rho(\bm r')} }_{\rho = \Phi}
\label{eq:delta_pi}
\end{equation}
and 
\begin{equation}
F^{(2)} := - \nabla\cdot \bm{\mathcal J}^{(2)}.
\end{equation}
Since $\rho^{(2)}=(\Frak t-\Frak t_0)F^{(2)}$, the quantity $\delta\bm\pi^{(1)}[\Phi]$ is the directional functional variation of $\bm\pi^{(1)}$ along the second-order density flow $F^{(2)}$.
Solving this equation and neglecting its exponentially decaying part, we obtain
\begin{align}
\bm p^{(3)} 
&= (\Frak t - \Frak t_0 - 1)\delta \bm \pi^{(1)}[\Phi] 
+\bm \pi^{(2)}[\Phi, \bm \pi^{(1)}] 
+ \bm \pi^{(3)}[\Phi]  \notag\\
&=: (\Frak t - \Frak t_0) \, \delta \bm \pi^{(1)}[\Phi] + \bm{\Frak p}^{(3)}[\Phi],
\end{align}
where
\begin{equation}
\bm{\Frak p}^{(3)}[\Phi] := - \delta \bm \pi^{(1)}[\Phi] +\bm \pi^{(2)}[\Phi, \bm \pi^{(1)}]
+ \bm \pi^{(3)}[\Phi] 
\end{equation}
is the non-secular part of $\bm p^{(3)}$.

\vspace{5mm}
\noindent \underline{$O(\epsilon^4)$:}
At fourth order, the time-independent part of the density current has contributions from $\bm J^{(1)}$, $\bm J^{(3)}$, and $\bm J^{(4)}$.
In addition, the second-order change in the density generates a contribution that grows linearly with $\Frak t-\Frak t_0$.
Collecting these terms gives
\begin{equation}
\p_{\Frak t} \rho^{(4)} = - \nabla\cdot\qty[
 \bm J^{(1)}[\Phi, \bm{\Frak p}^{(3)}] + \bm J^{(3)}[\Phi, \bm p^{(1)} ]+ \bm J^{(4)}[\Phi]
 + (\Frak t - \Frak t_0)\, \delta \bm{\mathcal J}^{(2)}[\Phi]
],
\end{equation}
where
\begin{align}
\delta \bm{\mathcal J}^{(2)}[\Phi] 
&:= 
\int_{V}\dd[2]\bm r'\ F^{(2)}(\bm r')\eval {\fdv{\bm{\mathcal J}^{(2)}[\rho](\bm r)}{\rho(\bm r')} }_{\rho = \Phi} \notag \\
&= \delta_\Phi \bm J^{(1)}[\Phi,\bm \pi^{(1)}] +   \bm J^{(1)}[\Phi, \delta\bm \pi^{(1)}] 
+ \delta \bm J^{(2)}[\Phi].
\end{align}
The first term in the second line is the variation of $\bm J^{(1)}$ with respect to density at fixed polarization, defined by
\begin{equation}
\delta_\Phi \bm J^{(1)}[\Phi,\bm \pi^{(1)}] 
:=
\int_{V}\dd[2]\bm r'\ F^{(2)}(\bm r')\eval {\fdv{\bm J^{(1)}[\rho, \bm p](\bm r)}{\rho(\bm r')} }_{\rho = \Phi,\ \bm p = \bm \pi^{(1)} }.
\end{equation}
The solution is given by 
\begin{equation}
\rho^{(4)} = -\nabla\cdot\qty[(\Frak t - \Frak t_0) \, \bm{\mathcal J}^{(4)} + \frac{1}{2}(\Frak t - \Frak t_0)^2 \delta \bm{\mathcal J}^{(2)}],
\end{equation}
where
\begin{equation}
 \bm{\mathcal J}^{(4)} := \bm J^{(1)}[\Phi, \bm{\Frak p}^{(3)}] + \bm J^{(3)}[\Phi, \bm \pi^{(1)} ]+ \bm J^{(4)}[\Phi].
\end{equation}
As at second order, a time-independent integration field has been absorbed into $\Phi$.
Combining $\rho^{(0)}$, $\epsilon^2\rho^{(2)}$, and $\epsilon^4\rho^{(4)}$, we reach Eq.~\eqref{eq:perturbation_solution}.

\section{Derivation of Equation~\eqref{eq:semigroup}}
\label{ape:semigroup}
In this Appendix, we derive the semigroup property, Eq.~\eqref{eq:semigroup}.
For simplicity of notation, we here denote $F^{(n)}[\Phi] = - \nabla\cdot \bm{\mathcal J}^{(n)}[\Phi]$.
We directly calculate the left-hand side of Eq.~\eqref{eq:semigroup}:
\begin{align}
\hat{\mathcal R}_{\Frak t''} \circ \hat{\mathcal R}_{\Frak t'}[\Phi] 
&= 
\hat{\mathcal R}_{\Frak t''}[\Phi'] \notag \\
&=
\Phi' + \qty (\epsilon^2 F^{(2)}[\Phi'] 
 + \epsilon^4 F^{(4)}[\Phi'])\, \Frak t'' 
 + \frac{\epsilon^4}{2}\delta F^{(2)}[\Phi']\, \Frak t''^2.
\end{align}
Noting that 
\begin{align}
F^{(2)}[\Phi'] 
&= 
F^{(2)}\qty[\Phi + \epsilon ^2 F^{(2)}[\Phi]\Frak t' + O(\epsilon^4)] \notag \\
&= 
F^{(2)}\qty[\Phi] + \epsilon^2 \Frak t' \, \delta F^{(2)}[\Phi] + O(\epsilon^4),
\end{align}
we obtain
\begin{align}
\hat{\mathcal R}_{\Frak t''} \circ \hat{\mathcal R}_{\Frak t'}[\Phi] 
&=
\Phi + \qty (\epsilon^2 F^{(2)}[\Phi] 
 + \epsilon^4 F^{(4)}[\Phi ])\qty( \Frak t' + \Frak t'')
 + \frac{\epsilon^4}{2}\delta F^{(2)}[\Phi] \qty( \Frak t'^2 + \Frak t''^2 + 2\Frak t' \Frak t'') + O(\epsilon^6) \notag \\
&=\Phi + \qty (\epsilon^2 F^{(2)}[\Phi] 
 + \epsilon^4 F^{(4)}[\Phi ])\qty( \Frak t' + \Frak t'')
 + \frac{\epsilon^4}{2}\delta F^{(2)}[\Phi] \qty( \Frak t' + \Frak t'')^2 + O(\epsilon^6).
\end{align}
Thus, Eq.~\eqref{eq:semigroup} holds up to the order $O(\epsilon^4)$.

\section{Invariance condition}
\label{ape:invariance}
Here, we show that $\bar{\bm p}[\Phi]$ satisfies the invariance condition, Eq.~\eqref{eq:invariance}.
From Eq.~\eqref{eq:hydrodynamics_ABP}, the left-hand side of Eq.~\eqref{eq:invariance} is 
\begin{equation}
\eval{\p_{\Frak t}\bm p}_{\rho=\Phi, \, \bm p=\bar{\bm p}[\Phi]}
= -\bar{\bm p}[\Phi] + \epsilon \bm \pi^{(1)}[\Phi] + \epsilon^2 \bm \pi^{(2)}[\Phi,\bar{\bm p}[\Phi]] +  \epsilon^3 \bm \pi^{(3)}[\Phi].
\end{equation}
Using $\bm {\pi}^{(2)}[\Phi,\bar{\bm p}[\Phi]] =\epsilon \bm \pi^{(2)}[\Phi,\bm {\pi}^{(1)}[\Phi]] + O(\epsilon^3)$, we have
\begin{equation}
\eval{\p_{\Frak t}\bm p}_{\rho=\Phi, \, \bm p=\bar{\bm p}[\Phi]} = \epsilon^3 \delta\bm {\pi}^{(1)}[\Phi] + O(\epsilon^5).
\end{equation}
We next calculate the right-hand side of Eq.~\eqref{eq:invariance}.
Using $\p_\Frak {t}\Phi = \epsilon^2 F^{(2)} + O(\epsilon^4)$ with $F^{(2)} = -\nabla \cdot \bm {\mathcal J}^{(2)}$ and $\bar{\bm p}[\Phi] = \epsilon \bm \pi^{(1)}[\Phi] + O(\epsilon^3)$, we obtain
\begin{align}
\eval{\p_{\Frak t}\bar{\bm p}[\rho]}_{\rho=\Phi, \, \bm p=\bar{\bm p}[\Phi]}
&=
\int_{V}\dd[2]\bm r'
\eval{\fdv{\bar{\bm p}[\rho](\bm r)}{\rho(\bm r')} }_{\rho=\Phi}\p_{\Frak t}\Phi (\bm r') \notag \\
&= 
\epsilon^3\int_{V}\dd[2]\bm r'\ F^{(2)}(\bm r')
\eval{\fdv{{\bm \pi}^{(1)}[\rho](\bm r)}{\rho(\bm r')} }_{\rho=\Phi} + O(\epsilon^5).
\end{align}
The integral in the last line equals
$\delta\bm{\pi}^{(1)}[\Phi]$, as defined in
Eq.~\eqref{eq:delta_pi}.
Thus, Eq.~\eqref{eq:invariance} holds up to $O(\epsilon^3)$.
If we use the adiabatic solution, Eq.~\eqref{eq:adiabatic}, the left-hand side of Eq.~\eqref{eq:invariance} is $O(\epsilon^5)$, whereas the right-hand side is $\epsilon^3\delta\bm {\pi}^{(1)}[\Phi] + O(\epsilon^5)$, and thus Eq.~\eqref{eq:invariance} does not hold. 

\section{Coefficients in the scalar field theories}
\label{ape:coefficients}
\subsection{Active Brownian particles}
\label{ape:coefficients_ABP}
The explicit expressions of the coefficients in Eq.~\eqref{eq:current_fourth_order} are given by
\begin{align}
\mathcal A(\Phi)
&=
u(\Phi)
\Bigg[
h(\Phi)\mathcal D'(\Phi)
-\mathcal D(\Phi)
+\frac{\eta(\Phi)}{2}
-b_0h(\Phi)
+\frac{c}{2}
-\frac{b_2}{32}
\Bigg]
\notag 
\\
&\ \ \ 
+
\frac{b_2}{16}u(\Phi)h(\Phi)
+c\,h(\Phi)
+\frac{b_2^2}{32}\Phi ,
\\
\mathcal B(\Phi)
&=
u(\Phi)
\Bigg[
h(\Phi)\mathcal D'(\Phi)
+\frac{u(\Phi)h(\Phi)}{8}
+\frac{\eta(\Phi)}{2}
-\left(
\frac{b_0}{2}+\frac{b_2}{4}
\right)h(\Phi)
+\frac{c}{2}
-\frac{b_2}{32}(1-2\Phi)
\Bigg] 
\notag\\
&\ \ \ 
-
\frac{b_2}{16}u(\Phi)h(\Phi)
+c\,h(\Phi)
-\frac{b_2^2}{32}\Phi  ,
\\
\mathcal C(\Phi)
&=
u(\Phi)
\Bigg[
h(\Phi)\mathcal D''(\Phi)
-\mathcal D'(\Phi)
+b_0
+\frac{9b_2}{32}
-\frac{3u(\Phi)}{16}
-\frac{h(\Phi)}{16}
\Bigg]
\notag \\
&\ \ \ +
\frac{b_2}{16}
\left[
h(\Phi)+u(\Phi)
\right]
-\frac{b_2^2}{32} ,
\\
\mathcal K(\Phi)
&=
k\Phi +
u(\Phi)h(\Phi)
\qty[\mathcal D(\Phi)
-\frac{\eta(\Phi)}{2}]
+\frac{3c}{2}u(\Phi)\Phi
-\frac{b_2}{32}u(\Phi)^2\Phi . \label{eq:stiffness}
\end{align}

\subsection{Chiral active Brownian particles}
\label{ape:coefficients_cABP}
Here, we write the tensor $\eta_{\alpha,\beta}(\rho)$ in Eq.~\eqref{eq:coefficients_hydro_cABP} as $\eta_{\alpha,\beta}(\rho) = \eta(\rho) \delta_{\alpha,\beta} + \eta_\perp(\rho) \varepsilon_{\alpha,\beta}$.
The coefficients in the parallel components are given by
\begin{align}
\mathcal A(\Phi)
&= 
\frac{u(\Phi)}{(1+\omega^2)^2}
\Bigg\{ 
(1-\omega^2)
\Bigg[
h(\Phi)\mathcal D'(\Phi)
-\mathcal D(\Phi)
-b_0h(\Phi)
+\frac{\eta(\Phi)}{2}
\Bigg]
+\omega\eta_\perp(\Phi)
\Bigg\}
\notag\\
&\ \ \ 
+\frac{c\qty[u(\Phi)+2h(\Phi)]}{2+2\omega^2}
-\frac{b_2\Phi u(\Phi)(2-\omega^2)}{(32+8\omega^2)(1+\omega^2)}
+\frac{b_2^2\Phi}{32+8\omega^2}, 
\\
\mathcal B(\Phi)
&= 
\frac{u(\Phi)}{(1+\omega^2)^2}
\Bigg\{
(1-\omega^2)
\Bigg[
h(\Phi)\mathcal D'(\Phi)
-\frac{b_0h(\Phi)}{2}
+\frac{\eta(\Phi)}{2}
+\frac{u(\Phi)h(\Phi)}{2(4+\omega^2)}
\Bigg]
\notag\\
&
\ \ \ +
\omega\qty[\eta_\perp(\Phi)-\frac{\omega u(\Phi)h(\Phi)}{2(4+\omega^2)}]
\Bigg\}
+\frac{c\qty[u(\Phi)+2h(\Phi)]}{2+2\omega^2}
-\frac{3b_2u(\Phi)h(\Phi)}{(8+2\omega^2)(1+\omega^2)}
\notag\\
&
\ \ \ 
-\frac{b_2^2\Phi}{32+8\omega^2},
\\
\mathcal C(\Phi)
&= 
\frac{u(\Phi)}{(1+\omega^2)^2}
\Bigg\{
(1-\omega^2)
\Bigg[
h(\Phi)\mathcal D''(\Phi)
-\mathcal D'(\Phi)
+b_0
-\frac{3u(\Phi)+h(\Phi)}{4(4+\omega^2)}
\Bigg]
\notag\\
&
\ \ \ 
+\frac{\omega^2\qty[3u(\Phi)+h(\Phi)]}{4(4+\omega^2)}
\Bigg\}
+\frac{b_2u(\Phi)}{4+4\omega^2}
+\frac{b_2\qty[3u(\Phi)+2h(\Phi)](2-\omega^2)}{(64+16\omega^2)(1+\omega^2)}
-\frac{b_2^2}{32+8\omega^2},
\\
\mathcal K(\Phi)
&= k\Phi
+\frac{u(\Phi)h(\Phi)}{(1+\omega^2)^2}
\Bigg\{
(1-\omega^2)
\qty[\mathcal D(\Phi)-\frac{\eta(\Phi)}{2}]
-\omega\eta_\perp(\Phi)
\Bigg\}
+\frac{3c\Phi u(\Phi)}{2+2\omega^2}
\notag\\
&
\ \ \ 
-\frac{b_2\Phi u^2(\Phi)(2-\omega^2)}{(64+16\omega^2)(1+\omega^2)}.
\end{align}
The coefficients in the perpendicular components are given by
\begin{align}
\mathcal A_\perp(\Phi)
&= 
\frac{u(\Phi)}{(1+\omega^2)^2}
\Bigg\{
-2\omega
\Bigg[
h(\Phi)\mathcal D'(\Phi)
-\mathcal D(\Phi)
-b_0h(\Phi)
+\frac{\eta(\Phi)}{2}
\Bigg]
+\frac{1-\omega^2}{2}\eta_\perp(\Phi)
\Bigg\}
\notag\\
&
\ \ \ 
-\frac{c\omega\qty[u(\Phi)+6h(\Phi)]}{2+2\omega^2}
+\frac{3b_2\omega\Phi u(\Phi)}{(32+8\omega^2)(1+\omega^2)}
-\frac{b_2^2\omega\Phi}{64+16\omega^2}, \label{eq:A_perp}
\\
\mathcal B_\perp(\Phi)
&= 
\frac{u(\Phi)}{(1+\omega^2)^2}
\Bigg\{
-2\omega
\Bigg[
h(\Phi)\mathcal D'(\Phi)
-\frac{b_0h(\Phi)}{2}
+\frac{\eta(\Phi)}{2}
+\frac{u(\Phi)h(\Phi)}{2(4+\omega^2)}
\Bigg]
\notag\\
&
\ \ \ 
+\frac{1-\omega^2}{2}
\qty[\eta_\perp(\Phi)-\frac{\omega u(\Phi)h(\Phi)}{2(4+\omega^2)}]
\Bigg\}
+\frac{c\omega\qty[2h(\Phi)-u(\Phi)]}{2+2\omega^2}
+\frac{3b_2\omega u(\Phi)h(\Phi)}{(16+4\omega^2)(1+\omega^2)}
\notag\\
&
\ \ \ 
+\frac{b_2^2\omega\Phi}{64+16\omega^2},
\\
\mathcal C_\perp(\Phi)
&= \frac{u(\Phi)}{(1+\omega^2)^2}
\Bigg\{
-2\omega
\Bigg[
h(\Phi)\mathcal D''(\Phi)
-\mathcal D'(\Phi)
+b_0
-\frac{3u(\Phi)+h(\Phi)}{4(4+\omega^2)}
\Bigg]
\notag\\
&
\ \ \ 
+\frac{\omega(1-\omega^2)\qty[3u(\Phi)+h(\Phi)]}{8(4+\omega^2)}
\Bigg\}
-\frac{3b_2\omega\qty[3u(\Phi)+2h(\Phi)]}{(64+16\omega^2)(1+\omega^2)}
+\frac{b_2^2\omega}{64+16\omega^2}, \label{eq:C_perp}
\\
\mathcal K_\perp(\Phi)
&= \frac{u(\Phi)h(\Phi)}{(1+\omega^2)^2}
\Bigg\{
-2\omega
\qty[\mathcal D(\Phi)-\frac{\eta(\Phi)}{2}]
-\frac{1-\omega^2}{2}\eta_\perp(\Phi)
\Bigg\}
-\frac{3c\omega\Phi u(\Phi)}{2+2\omega^2}
\notag\\
&
\ \ \ 
+\frac{3b_2\omega\Phi u^2(\Phi)}{(64+16\omega^2)(1+\omega^2)}.
\end{align}

\section{Field theory in physical units}
\label{ape:physical_units}
In the main text, we work with dimensionless variables scaled by $\ell_\RM{macro}$, $\tau_\RM{macro}$, and the reference density $\rho_* = 1/\gamma$. 
Here, we summarize the field equations obtained in Sections~\ref{subsec:field_theory_ABP} and \ref{subsec:field_theory_cABP} in the original physical units.

To return to the original physical units, we make the replacements
\begin{align}
\begin{split}
\bm r &\to \frac{\bm r}{\ell_\RM{macro}},
\ \ \ \ \ \ \ \ \ \ 
t \to \frac{t}{\tau_\RM{macro}}, \\
\Phi &\to \frac{\Phi}{\rho_*},
\ \ \ \ \ \ \ \ \ \ \ \ \ 
\bm J \to \frac{\tau_\RM{macro}}{\rho_*\ell_\RM{macro}}\bm J.
\end{split}
\end{align}
Recall that in Section~\ref{subsec:scaling} we set $\epsilon = \epsilon_\ell = \sqrt{\epsilon_\tau}$, meaning that $\ell_\RM{macro}^2/\tau_\RM{macro} = \ell_\RM{p}^2/\tau_\RM{p}$.
Using this relation, the total current $\bm J$ for ABPs in physical units reads
\begin{align}
\bm J
=
-\hat{\mathcal D}(\Phi)\nabla\Phi
+\hat{\mathcal K}(\Phi)\nabla\nabla^2\Phi
+\hat{\mathcal A}(\Phi)(\nabla^2\Phi)\nabla\Phi +\hat{\mathcal B}(\Phi)\nabla\abs{\nabla\Phi}^2
+\hat{\mathcal C}(\Phi)\abs{\nabla\Phi}^2\nabla\Phi,
\end{align}
with the dimensional coefficients given by
\begin{align}
\begin{split}
&\hat{\mathcal D}(\Phi)
=\frac{\ell_\RM{p}^2}{\tau_\RM{p}}\mathcal D\qty(\frac{\Phi}{\rho_*}),
\ \ \ \ \ \ \ \ \ 
\hat{\mathcal K}(\Phi)
=\frac{\ell_\RM{p}^4}{\tau_\RM{p}}\mathcal K\qty(\frac{\Phi}{\rho_*}),
\\
&\hat{\mathcal A}(\Phi)
=\frac{\ell_\RM{p}^4}{\tau_\RM{p}\rho_*}\mathcal A\qty(\frac{\Phi}{\rho_*}),
\ \ \ \ \ \ 
\hat{\mathcal B}(\Phi)
=\frac{\ell_\RM{p}^4}{\tau_\RM{p}\rho_*}\mathcal B\qty(\frac{\Phi}{\rho_*}),
\\
&\hat{\mathcal C}(\Phi)
=\frac{\ell_\RM{p}^4}{\tau_\RM{p}\rho_*^2}\mathcal C\qty(\frac{\Phi}{\rho_*}),
\end{split}
\end{align}
where $\mathcal D$ is given by Eq.~\eqref{eq:diffusion_coeff_ABP}, and $\mathcal K(\Phi)$, $\mathcal A(\Phi)$, $\mathcal B(\Phi)$, and $\mathcal C(\Phi)$ are given in Appendix~\ref{ape:coefficients_ABP}. 
In the same way as in Section~\ref{subsec:field_theory_ABP}, the total current can also be written in the form
\begin{align}
\bm J
=
-\nabla\fdv{\mathcal F[\Phi]}{\Phi}
-\hat{\lambda}(\Phi)\nabla\abs{\nabla\Phi}^2 
+\hat{\zeta}(\Phi)(\nabla^2\Phi)\nabla\Phi
+\hat{\nu}(\Phi)\abs{\nabla\Phi}^2\nabla\Phi
\end{align}
with
\begin{align}
\begin{split}
\hat{\lambda}(\Phi)
&=\frac{\ell_\RM{p}^4}{\tau_\RM{p}\rho_*}
\qty[\frac{1}{2}\mathcal K'\qty(\frac{\Phi}{\rho_*})-\mathcal B\qty(\frac{\Phi}{\rho_*})], 
\ \ \ \ \ \ 
\hat{\zeta}(\Phi)
=\frac{\ell_\RM{p}^4}{\tau_\RM{p}\rho_*}
\qty[\mathcal A\qty(\frac{\Phi}{\rho_*})-\mathcal K'\qty(\frac{\Phi}{\rho_*})], \\
\hat{\nu}(\Phi)
&=\frac{\ell_\RM{p}^4}{\tau_\RM{p}\rho_*^2}
\qty[\mathcal C\qty(\frac{\Phi}{\rho_*})-\frac{1}{2}\mathcal K''\qty(\frac{\Phi}{\rho_*})].
\end{split}
\end{align}
Note that the coefficients in physical units do not depend on either $\ell_\RM{macro}$ or $\tau_\RM{macro}$, as expected. 
We can therefore, in principle, determine the values of these coefficients using the microscopic parameters, interaction potential, and pair distribution function as inputs.

One can also write the field equation for the cABP model in the original physical units in the same way:
\begin{align}
\p_t\Phi
=
-\nabla\cdot\bm J^\parallel
-\hat{\zeta}_\perp(\Phi)\qty{\nabla^2\Phi,\Phi}
-\hat{\nu}_\perp(\Phi)\qty{\abs{\nabla\Phi}^2,\Phi}
\end{align}
with
\begin{align}
\hat{\zeta}_\perp(\Phi)
=\frac{\ell_\RM{p}^4}{\tau_\RM{p}\rho_*}
\qty[\mathcal A_\perp\qty(\frac{\Phi}{\rho_*})-\mathcal K'_\perp\qty(\frac{\Phi}{\rho_*})], 
\ \ \ 
\hat{\nu}_\perp(\Phi)
=\frac{\ell_\RM{p}^4}{\tau_\RM{p}\rho_*^2}
\qty[\mathcal C_\perp\qty(\frac{\Phi}{\rho_*})-\mathcal B'_\perp\qty(\frac{\Phi}{\rho_*})],
\end{align}
where $\mathcal A_\perp(\Phi)$ and $\mathcal C_\perp(\Phi)$ are given by Eqs.~\eqref{eq:A_perp} and \eqref{eq:C_perp}, respectively.
\newcommand{\newblock}{}
\nocite{apsrev42Control}
\bibliographystyle{apsrev4-2}
\bibliography{ref}
\end{document}